\documentclass[]{aa}
\usepackage[varg]{txfonts}
\usepackage{graphicx}
\usepackage{amsmath}
\usepackage{amssymb}
\usepackage{footmisc}
\usepackage{lscape}
\usepackage{color}
\usepackage{longtable}
\usepackage{natbib,twoopt}

\newcommand{\hi}{\mbox{H\,{\sc i}}} 

\begin{document} 

\title{Spectrophotometric analysis of GRB afterglow extinction curves with X-shooter\thanks{Based on observations collected at the European Southern Observatory, Paranal, Chile, under programs 084.A-0260(B), 085.A-0009(B), 088.A-0051(B), 089.A-0067(B) and 091.C-0934(C).}}

\author{
J. Japelj\inst{1}
\and
S. Covino\inst{2} 
\and 
A. Gomboc\inst{1}
\and 
S. D. Vergani\inst{3,2}
\and 
P. Goldoni\inst{4}
\and
J. Selsing\inst{5}
\and
Z. Cano\inst{6}
\and
V. D'Elia\inst{7,8}
\and
H. Flores\inst{3}
\and
J. P. U. Fynbo\inst{5}
\and
F. Hammer\inst{3}
\and
J. Hjorth\inst{5}
\and
P. Jakobsson\inst{6}
\and
L. Kaper\inst{9}
\and
D. Kopa\v{c}\inst{10}
\and
T. Kr\"{u}hler\inst{11,5}
\and
A. Melandri\inst{2}
\and
S. Piranomonte\inst{7}
\and
R.~S\'{a}nchez-Ram\'{\i}rez\inst{12,13,14}
\and
G. Tagliaferri\inst{2}
\and
N. R. Tanvir\inst{15}
\and
A. de Ugarte Postigo\inst{14}
\and
D. Watson\inst{5}
\and
R. A. M. J. Wijers\inst{9}
}
  
\institute{Faculty of Mathematics and Physics, University of Ljubljana, Jadranska ulica 19, SI-1000 Ljubljana, Slovenia \newline\email{jure.japelj@fmf.uni-lj.si} 
\and INAF, Osservatorio Astronomico di Brera, via E. Bianchi 46, 23807 Merate, Italy
\and GEPI-Observatoire de Paris Meudon. 5 Place Jules Jannsen, F-92195, Meudon, France
\and APC, Astroparticule et Cosmologie, Univ. Paris Diderot, CNRS/IN2P3, CEA/Irfu, Obs. de Paris, Sorbonne Paris Cité, 10 rue Alice Domon et Léonie Duquet, 75205 Paris Cedex 13, France
\and Dark Cosmology Centre, Niels Bohr Institute, University of Copenhagen, Juliane Maries Vej 30, 2100 Copenhagen Ø, Denmark
\and Centre for Astrophysics and Cosmology, Science Institute, University of Iceland, Dunhagi 5, 107 Reykjavík, Iceland
\and INAF-Osservatorio Astronomico di Roma, Via di Frascati 33, 00040 Monte Porzio Catone, Italy
\and ASI-Science Data Center, Via del Politecnico snc, I-00133 Rome, Italy
\and Astronomical Institute Anton Pannekoek, University of Amsterdam, PO Box 94249, 1090 GE Amsterdam, the Netherlands
\and ARI – Liverpool John Moores University, IC2 Liverpool Science Park, 146 Brownlow Hill, Liverpool, L3 5RF, UK
\and European Southern Observatory, Alonso de Co\'{o}rdova 3107, Vitacura, Casilla 19001, Santiago 19, Chile
\and Unidad Asociada Grupo Ciencias Planetarias (UPV/EHU, IAA-CSIC), Departamento de F\'{\i}sica Aplicada I, \\ E.T.S. Ingenier\'{\i}a, Universidad del Pa\'{\i}s Vasco (UPV/EHU), Alameda de Urquijo s/n, E-48013 Bilbao, Spain.
\and Ikerbasque, Basque Foundation for Science, Alameda de Urquijo 36-5, E-48008 Bilbao, Spain.
\and Instituto de Astrof\'{\i}sica de Andaluc\'{\i}a (IAA-CSIC), Glorieta de la Astronom\'{\i}a s/n, E-18008, Granada, Spain.
\and Department of Physics and Astronomy, University of Leicester, University Road, Leicester LE1 7RH, UK
}
 

\date{Received DD Mmmm YYYY / Accepted DD Mmmm YYYY} 
   
\abstract{In this work we use gamma-ray burst (GRB) afterglow spectra observed with the VLT/X-shooter spectrograph to measure rest-frame extinction in GRB lines-of-sight by modeling the broadband near-infrared (NIR) to X-ray afterglow spectral energy distributions (SEDs). Our sample consists of nine {\it Swift} GRBs, eight of them belonging to the long-duration and one to the short-duration class. Dust is modeled using the average extinction curves of the Milky Way and the two Magellanic Clouds. We derive the rest-frame extinction of the entire sample, which fall in the range $0 \lesssim {\it A}_{\rm V} \lesssim 1.2$. Moreover, the SMC extinction curve is the preferred extinction curve template for the majority of our sample, a result which is in agreement with those commonly observed in GRB lines-of-sights. In one analysed case (GRB\,120119A), the common extinction curve templates fail to reproduce the observed extinction. To illustrate the advantage of using the high-quality X-shooter afterglow SEDs over the photometric SEDs, we repeat the modeling using the broadband SEDs with the NIR-to-UV photometric measurements instead of the spectra. The main result is that the spectroscopic data, thanks to a combination of excellent resolution and coverage of the blue part of the SED, are more successful in constraining the extinction curves and therefore the dust properties in GRB hosts with respect to photometric measurements. In all cases but one the extinction curve of one template is preferred over the others. We show that the modeled values of the extinction $A_{\rm V}$ and the spectral slope, obtained through spectroscopic and photometric SED analysis, can differ significantly for individual events, though no apparent trend in the differences is observed. Finally we stress that, regardless of the resolution of the optical-to-NIR data, the SED modeling gives reliable results only when the fit is performed on a SED covering a broader spectral region (in our case extending to X-rays).}
 
\keywords{Gamma-ray bursts: general - ISM: dust,extinction}
 
\authorrunning{J. Japelj et al.} 
\titlerunning{GRB afterglow extinction curves with X-shooter}
\maketitle
  
\section{Introduction}
Dust plays a central role in the astrophysical processes of interstellar medium and in the formation of stars \citep[e.g.,][]{Mathis1990,Draine2003}. Its obscuring effects can introduce large uncertainties to the interpretation of astronomical observations, but at the same time offer us the means to study its physical properties, e.g.: the attenuation of light as a function of wavelength, or extinction curve, is strongly dependent on the composition and size distribution of the dust grains. The origin and properties of dust are still poorly known, especially at cosmological distances, where a unique probe of dust can be found in gamma-ray bursts (GRBs).

GRBs originate in galaxies at cosmological distances \citep[e.g.,][]{Bloom1998,Jakobsson2012}. As they are usually accompanied by bright optical and X-ray afterglow emission \citep[e.g.,][]{Kann2011}, they can be used as a powerful tool to study environments at different stages of the Universe's evolution. According to the standard theory \citep[e.g.,][]{Gehrels2009,Gomboc2012}, GRB afterglows are a result of an interaction between highly relativistic ejecta, produced in the progenitor's explosion, and an interstellar medium. The resulting emission is of a synchrotron nature, which in its simplest case, has a power-law dependence  in both time and frequency \citep[e.g.,][]{Sari1998}. The deviation from a simple power-law is then attributed to the absorption and scattering of light by dust grains \citep[e.g.,][]{Kann2006,Kann2011}. GRB afterglow emission is thus better suited for studying the extinction, as compared to more complex spectra of quasars or galaxies. Modeling of afterglow spectral energy distribution (SED) in optical--to--X-ray spectral range provides us with information regarding the intrinsic afterglow spectrum and the properties of the intervening dust. The dust properties in random lines-of-sight (LOS) in high-redshift galaxies are not known \textit{a priori}, therefore representative extinction curves are usually assumed in the modeling. Best studied and widely adopted are the average extinction curves observed in the Milky Way \citep[MW; ][]{Cardelli1989,Pei1992,Fitzpatrick2007}, Large Magellanic Cloud \citep[LMC; ][]{Pei1992,Gordon2003} and Small Magellanic Cloud \citep[SMC; ][]{Prevot1984,Pei1992,Gordon2003}. The three curves differ in their UV slope (steepest in the SMC and shallowest in MW curve) and the prominence of the 2175 $\AA$ bump feature, the latter being the strongest in the MW-type and disappearing in the SMC-type dust.

Afterglow SED modeling, either limited to the near-infrared to ultraviolet (NIR-UV) \citep[e.g.,][]{Galama2001,Stratta2004,Kann2006,Kann2010,Liang2010} or extended to the X-ray spectral range \citep{Schady2007,Schady2010,Greiner2011}, has revealed systems with mostly low LOS extinction where the extinction is preferentially described with the SMC-type dust. Studies dedicated to subsamples of more extincted afterglows \citep{Kruhler2011,Zafar2012,Perley2013,Fynbo2014} indicated that some SEDs show signatures of the 2175 $\mathrm{\AA}$ absorption feature. To get a clear and unbiased picture of the extinction properties, \citet{Covino2013} analysed the SEDs of a complete sample of GRBs, which were not biased towards optically bright events \citep{Salvaterra2012,Melandri2012}. They find that $\sim 50\%$ of the afterglows are found within lines-of sight of low extinction ($A_{\rm V} < 0.4$ mag) and only $\sim 13\%$ are heavily extincted ($A_{\rm V} > 2$ mag). 

Most of the extinction studies have been done using photometric SEDs. An accurate extinction measurement requires simultaneous high-quality data in a broad spectral range from NIR to X-rays. Examples of reliable studies of homogeneous data sets are those using the observations done with the GROND\footnote{http://www.mpe.mpg.de/$\sim$jcg/GROND/} instrument \citep{Greiner2008}, whose capability to simultaneously observe in 7 bands in the nIR-UV is especially well suited for SED studies \citep{Greiner2011,Kruhler2011}. 

Extinction studies could be improved by using spectroscopic instead of photometric SEDs. \citet{Zafar2011} (hereafter Z11) studied a sample of 41 optical afterglow spectra obtained mostly with FORS\footnote{http://www.eso.org/sci/facilities/paranal/instruments/fors.html} (VLT). They showed that the SEDs with prominent 2175 $\mathrm{\AA}$ absorption are among the most extincted with $A_{\rm V} > 1.0$, while the events with SMC-type dust have low extinction ($A_{\rm V} < 0.65$). This led them to conclude that the low detection rate of afterglows with LMC or MW-type dust is more likely due to observational bias against dusty LOS than due to the MW-type dust to be rare in high redshift environments. While of a limited spectral coverage, the spectroscopic SEDs allowed them to prove convincingly that the afterglow spectra are indeed consistent with a simple synchrotron model.

In this work we take the next step in the SED modeling by using the afterglow spectra obtained with the X-shooter instrument \citep{Vernet2011}. X-shooter is a state-of-the-art intermediate resolution spectrograph, mounted on the VLT, that simultaneously covers a broad spectral range with three spectroscopic arms: ultraviolet (UVB; $\sim$3000-5500 $\mathrm{\AA}$), visible (VIS; $\sim$5500-10000 $\mathrm{\AA}$) and near-infrared (NIR; $\sim$10000-25000 $\mathrm{\AA}$). The large spectral coverage of X-shooter spectra offers us a unique opportunity to apply a detailed extinction curve analysis. The power of X-shooter in extinction curve studies was illustrated by the recent observation of GRB\,140506A, whose spectrum revealed a unique extinction signature in an afterglow spectrum: a very strong flux drop below $\sim 4000 ~\mathrm{\AA}$ (in the GRB's rest system) is unprecedented in the study of GRB environment and has been found only in a few other LOS to other types of objects so far \citep{Fynbo2014}. Our aim is to use afterglow spectra acquired with the X-shooter instrument to derive the dust properties in the LOSs of GRB host galaxies and to evaluate the applicability of the commonly used extinction curves in this type of analysis. The sample, data preparation and analysis are presented in Section \ref{prepare}. Results and detailed discussion are given in Sections \ref{results} and \ref{discuss}, respectively. We summarize our conclusions in Section \ref{conclude}.

Throughout the paper the convention $F_{\nu}(t) \propto t^{-\alpha}\nu^{-\beta}$ is adopted, where $\alpha$ and $\beta$ are temporal and spectral afterglow slopes, respectively. Times are given with respect to GRB trigger time. 

\section{Data and Analysis}
\label{prepare}

\subsection{Preparation of X-shooter and X-ray spectra}

As a part of the X-shooter GRB GTO program\footnote{PIs: J. Fynbo and L. Kaper}, spectra for $\sim 60$ GRB afterglows have been acquired in the period between 2009 and 2014. From this sample we selected those GRBs whose spectra, according to photometric observations of the afterglows, are not contaminated by supernova or host galaxy emission. The spectra were reduced and calibrated using version 2.0 of the X-shooter data reduction pipeline \citep{Goldoni2006,Modigliani2010} - details are described in Fynbo et al. (in prep). In particular, the instrument's response function, required to flux-calibrate a spectrum, was obtained by comparing an observed spectrum of a spectrophotometric standard star with the tabulated values. Flux calibration of the X-shooter spectra has to be robust for our science case. The observations of the standard stars used for the flux calibration are performed only once per night with a $5''$ wide slit and binning different than the one used for the science spectra. Afterglow observations are done with much narrower slits, usually with $1.0''$, $0.9''$ and $0.9''$ for UVB, VIS and NIR spectra, respectively. Even if the sky conditions are the same when the standard star and afterglow are observed, a narrow slit loses much more light in the case where seeing is comparable to or larger than the slit width, which can be especially problematic due to the weak wavelength dependence of the seeing \citep{Roddier1981}. Other potential problems are the loss of light in the case the slit is not positioned in the direction of the parallactic angle, and a faulty performance of the atmospheric dispersion correctors in the UVB and VIS spectroscopic arms, which have been disabled in Aug 2012.

\begin{table}[htp]
\renewcommand{\arraystretch}{1.5}
\begin{center}
\small
\caption{Presentation of the sample}
\begin{tabular}{lccccr}
\hline\hline
\vspace{-0.15cm}
GRB & $z$ & $A_{\rm V}^{\rm G}$ & $N_{\rm H,X}^{\rm G}$                  & $T_{\rm mid}$ & $\Delta T_{\rm X}$\\
       &    &                                  & $[10^{20} {\rm cm}^{-2}]$   &       days  & ks\\
\hline
100219A & 4.667  &  0.208  &  6.7 & 0.55   & 15 - 50\\
100418A & 0.624  &  0.200  &  4.8 & 1.47	  & 80 - 250\\
100814A & 1.44    &  0.054  &  1.6 & 4.1     & 200 - 400\\
100901A & 1.408  &  0.270  &  7.3 & 2.75   & 160 - 500\\
120119A & 1.728  &  0.295  &  7.7 & 0.074 & 4 - 20\\
120815A & 2.358  &  0.320  &  8.4 & 0.086 & 6 - 20\\
130427A & 0.34	  &  0.055  &  1.8 & 0.70   & 40 - 70\\
130603B$^{(s)}$  & 0.357  & 0.063  &  2.0 & 0.35   &  15 - 60\\
130606A & 5.913  &  0.066  &  2.0 & 0.329 & 10 - 40\\
\hline
\end{tabular}
\tablefoot{GRBs in the sample. For each GRB we report the basic information required in the analysis: redshift, Galactic extinction $A_{\rm V}^{\rm G}$ and equivalent neutral hydrogen column density $N_{\rm H,X}^{\rm G}$ in the burst's line-of-sight, mid-time of the X-shooter observation $T_{\rm mid}$ and the time interval $\Delta T_{\rm X}$ (in ks) used for the construction of the X-ray part of the SED. Times are given relative to the start of the GRB $\gamma$-ray emission in the observer frame. GRB\,130603B belongs to the short class, while other GRBs in the sample are of the long class.}
\label{tabsample}
\end{center}
\end{table}

For these reasons we cannot blindly rely on the instrument's response function obtained in the calibration process and we therefore require multiwavelength photometric data of the studied objects to check and validate the flux calibration (see also Kr\"uhler et al. in prep. ). Our sample is thus limited by the availability of multi-band photometric data in the literature. Photometric observations should be available at or near the epoch of X-shooter spectra. We found that for several GRBs in the full X-shooter sample not only the absolute flux level, but also the flux calibration as a function of wavelength did not match the photometric SED. Since it is extremely difficult to reliably account and correct for all the effects (either of observational or technical nature) which influence the calibration, we decided to work only with the spectra with flux calibration where, after applying a correction to the absolute flux calibration, the difference between the spectrum and photometry is less than 10$\%$ in all bands. The photometric data are used only to check the validity of the flux calibration -- we do not use them to correct the slopes of the flux-calibrated spectra. UV photometric observations are seldom available around the X-shooter epoch. The seeing for all but one event, as measured from the 2D trace of the spectra, is small enough that the slit losses as a function of wavelength at the blue part of the UVB arm should be negligible and therefore we trust the calibration of the bluest SED part. The seeing was bad during the observation of the GRB\,100901A afterglow ($FWHM \sim 1.5^{\arcsec}$), but in this case the UV photometric observations were available, confirming the calibration in the UV part of the SED is fine. In summary, a GRB was included in our sample if:
\begin{enumerate}
\item multwavelength photometric observations of its afterglow around the epoch of X-shooter observation are available,
\item its spectrum is not contaminated by host or supernova emission,
\item the difference between the flux-calibrated spectrum and photometry is less than 10$\%$ in all bands. 
\end{enumerate}
Our final sample consists of nine GRBs, eight of them belonging to the long and one to the short class\footnote{Traditionally, GRBs are classified into a long or short class according to their observed duration (i.e., longer or shorter than $\sim 2$ s) and spectral properties \citep{Kouveliotou1993}. However, a reliable classification into the two classes, which correspond to different progenitor types, is usually more complicated \citep[e.g.,][]{Zhang2009}.}. The sample is presented in Table \ref{tabsample}.  In the future, more published light curve data will enable us to expand the analysis to a bigger sample.

The absolute flux calibration of the spectra in this sample was first fine-tuned with the photometric observations. Spectra were corrected for the extinction originating in our Galaxy using \citet{Cardelli1989} extinction curve (assuming the ratio of total-to-selective extinction $R_{\rm V} = 3.1$) and Galactic extinction maps \citep{Schlafly2011}. Regions of telluric absorption and strong absorption lines originating in the GRB host galaxies were masked out. Spectra were then rebinned in bins of widths 30 -- 100 $\mathrm{\AA}$ in order to reduce the noise and to guarantee a comparable weight of the optical and X-ray SED part. The binning was performed with the sigma-clipping algorithm: all data points within a binned interval that differ for more than three standard deviations from the mean of the points in the interval were rejected. Since the flux in the binning region is not constant (i.e., the spectra follow power-laws), we paid special attention to avoid removing the tails of the binned region during the sigma-clipping. Due to a large number of points that were binned we calculated the errors as an average of a standard deviation of binned points from the mean value $\sigma/\sqrt{N}$, where $N$ is the number of binned data points. We cross-checked these values with a Monte Carlo simulation. We assumed the errors of the data points, as obtained in the reduction and calibration procedure, are Gaussian. We then resampled the data points in each interval for a thousand times and computed the 1-$\sigma$ equivalent of the resulting distribution of mean values. The errors, computed by the two methods, are comparable.

The reddest part (e.g., $K$ band) of the NIR spectrum is seldom accurately flux calibrated. This is due mainly to the strong vignetting of the K band (section 2.4.9 of the X-shooter User Manual) which prevents a reliable sky subtraction especially for long exposure and faint sources. In addition, many observations were conducted with a special $K$-blocking filter\footnote{The filter blocks the spectral range above 2 $\mu$m and thus prevents the scattered light from $K$-band orders to contaminate the $J$- and $H$-band background \citep[e.g.,][]{Vernet2011}.}. For these reasons, only $\lambda \lesssim 20000 ~\mathrm{\AA}$ part of the NIR spectra have been used in the analysis.

X-ray data from the {\it Swift}/XRT instrument \citep{Burrows2005} were taken from the online repository of X-ray afterglow spectra \citep{Evans2009}. The X-ray SED was built from the light curve in a time interval around the epoch at which the X-shooter spectrum was taken: only a portion of the X-ray light curve without significant spectral evolution was considered. The mean time of the X-ray SED was computed as $\sum_{i}(t_{i}\Delta t_{i})/\sum_{i}(\Delta t_{i})$, where $t_{i}$ is the mid-time of individual exposure and $\Delta t_{i}$ is the exposure length. The light curve at the considered time interval was fitted with a power-law function: by knowing the slope, the X-ray SED was normalized by interpolation to the epoch of X-shooter observations. The uncertainty of calculated normalization is never greater $\sim 10\%$; by varying the normalization within this uncertainty for each analysed SED we found that the best fit parameters are consistent within errors. X-ray channels were rebinned in order to have at least 20 photons in each channel, which ensures that the data are roughly Gaussian and allows a reliable use of $\chi^{2}$ statistics.

\subsection{SED modeling}
\label{modeling}
In most cases the optical-to-X-ray SED of the afterglow is studied hours to days after the GRB explosion (see Table \ref{tabsample}), when the shape is expected to depend on the relative values of the cooling ($\nu_{\rm c}$), optical ($\nu_{\rm O}$), and X-ray ($\nu_{\rm X}$) frequencies \citep{Sari1998}. The SED can therefore be described with a power-law function $F_{\nu} = F_{0} \nu^{-\beta_{\rm OX}}$ (i.e., $\nu_{\rm c} < \nu_{\rm O}$ or $\nu_{\rm c} > \nu_{\rm X}$) or, in case $\nu_{\rm O} < \nu_{\rm c} < \nu_{\rm X}$, by a broken power-law:  
\begin{equation}
\label{eq:powerlaw}
F_{\nu} = F_{0} \left \lbrace
\begin{array}{ll}
\nu^{-\beta_{\rm O}} & \nu \leq \nu_{\rm c}\\
\nu_{\rm c}^{\Delta \beta} \nu^{-\beta_{\rm X}} & \nu > \nu_{\rm c},\\
\end{array}
\right.
\end{equation}
where $\Delta \beta = \beta_{\rm X} - \beta_{\rm O}$.

\begin{figure*}[t]
\centering
\begin{tabular}{cc}
\includegraphics[scale=0.40]{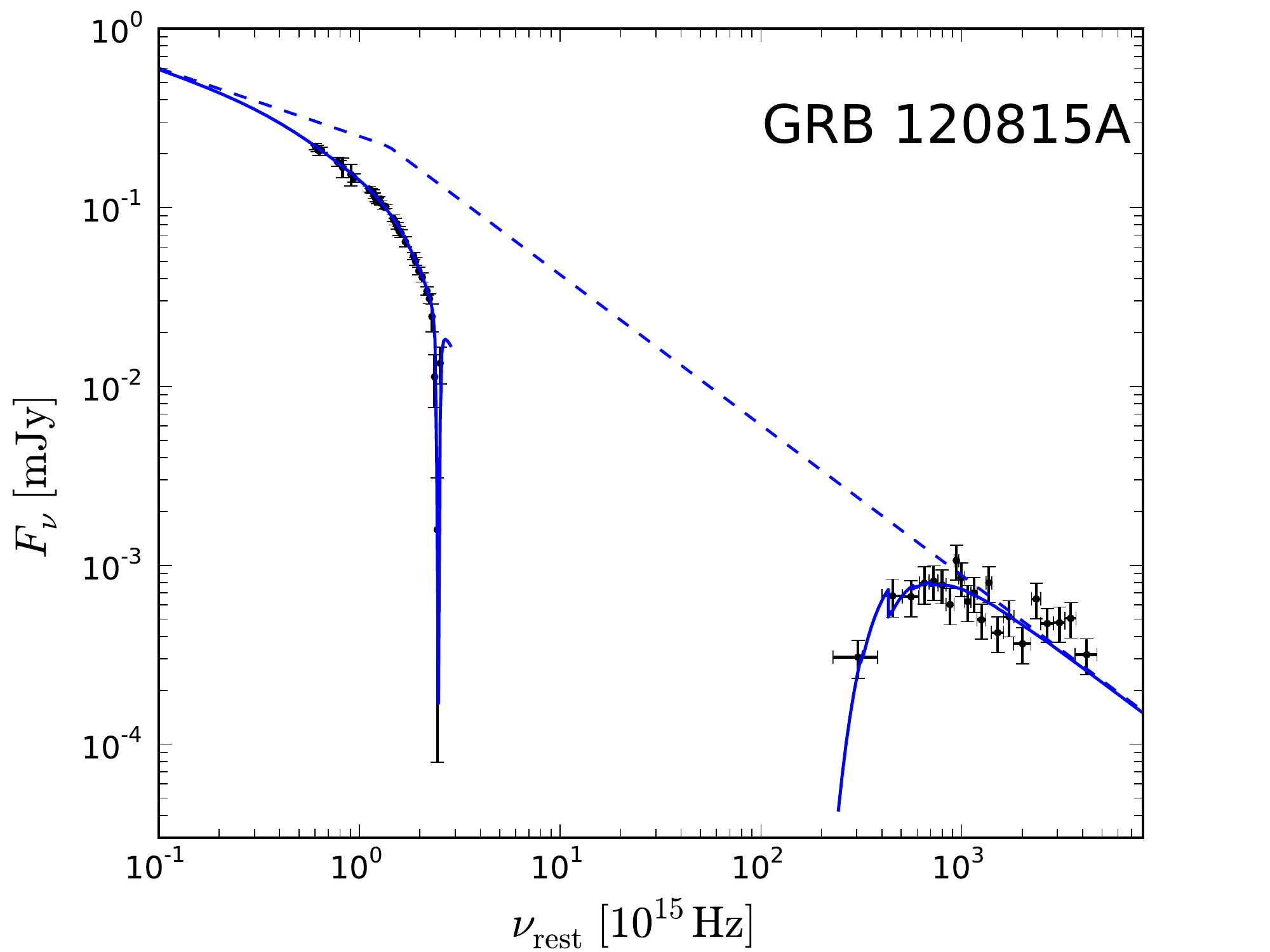}&
\includegraphics[scale=0.40]{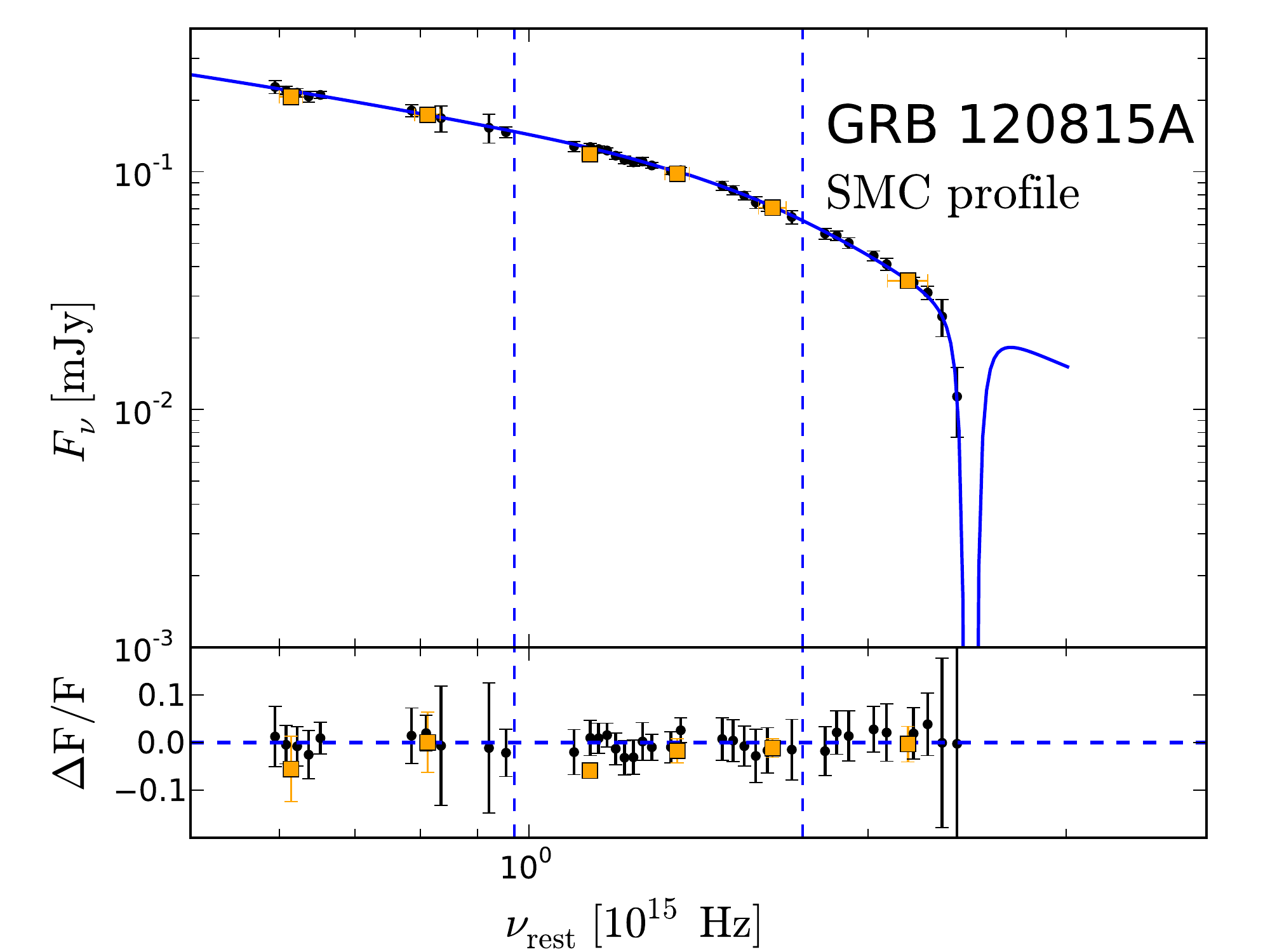}\\
 \end{tabular}
\caption{Rest-frame afterglow SED analysis of GRB 120815A. Broadband SED is shown in the left plot. Optical and X-ray data are plotted with black points, the best fitted model is plotted with solid blue lines and the intrinsic afterglow spectrum with dashed blue line. Zoomed part of the X-shooter SED is shown in the right plot, where residuals to the best-fitted model are also shown. Vertical lines divide the spectral regions covered by the NIR, VIS and UVB spectrograph arm. Orange points are photometric observations, used to calibrate the absolute flux of the spectrum. References for photometric data points and details on the fitting procedure are reported in Section \ref{results}. Similar plots for other GRBs in our sample are shown in Figures \ref{figsed1} and \ref{figsed2}.}
 \label{sed1}
\end{figure*}   

The optical afterglow is attenuated due to light scattering by dust particles and can be accounted for in our observations by: $F_{\nu,\mathrm{obs}} = F_{\nu} 10^{-0.4A_{\lambda} }$, where $A_{\lambda}$ is a wavelength-dependent extinction in the host galaxy frame. We assume the extinction is dominated by the dust in GRB hosts rather than being attenuated by intervening galaxies that happen to occur in the LOS between the GRB event and observers on Earth - there are known cases when other galaxies are found in the LOS \citep[e.g.,][]{Sparre2014}, however in most cases the host galaxy is the dominant absorption system \citep[e.g.,][]{Vergani2009,Schady2010}. The value of $A_{\lambda} = A_{\rm V}f(\lambda; R_{\rm V}, \ldots)$ is parametrised using the rest-frame extinction $A_{\rm V}$ (extinction in the rest frame $V$-band, at 5500 {\AA}) and the extinction curve $f$, which reflects the properties of dust in the line-of-sight. We consider the average observed extinction curves of Milky Way (MW) and Large and Small Magellanic Clouds \citep[LMC, SMC; ][]{Pei1992} in the analysis. Initially we also considered the average extinction curve of starburst galaxies \citep{Calzetti2000}, but we find the starburst curve is inadequate in describing extinction in lines-of-sight of GRB hosts - in addition to being statistically unjustified, fits of starburst models for all events in our sample result in physically unreasonable best-fit parameters of $\beta$ and $A_{\rm V}$. The conclusion that starburst models cannot describe dust in GRB lines-of-sight was also reached in other studies \citep[e.g.,][]{Covino2013}. This is not surprising as it is not expected that extinction in LOS to individual sources and galaxy integrated extinction properties are the same.

Fitting only the X-shooter SED may result in a degenerate solution between the values of the spectral slope, $\beta$, and the visual extinction, $A_{\rm V}$, i.e., data could be successfully represented both by a shallow slope and high amount of extinction or a steep slope and low extinction. The problem can be resolved by constraining one of the parameters. We decided to constrain the spectral slope by including the X-ray part of the SED in the fit. By analysing a large afterglow sample, Z11 found that the difference between the optical and X-ray slope is very close to the theoretically predicted value of $\Delta \beta = 0.5$ in the majority of cases. Nevertheless, we left the slope difference $\Delta \beta$ as a free parameter of our model. The photoelectric absorption of soft X-rays by metals is assumed to originate in the Galaxy and in the host galaxy at a known redshift $z$. Galactic equivalent neutral hydrogen column density $N_{\rm H,X}^{G}$ is taken from maps provided by \citet{Kalberla2005}\footnote{In general, in addition to the atomic hydrogen \hi, provided by maps of \citet{Kalberla2005}, a contribution of molecular hydrogen H$_{2}$ should be taken into account when estimating Galactic X-ray absorption \citep{Willingale2013}. We ignore the molecular component in order to make our modeling of the X-ray spectra comparable to analyses of larger samples in previous studies (e.g., see Section \ref{ext_and_hyd} and Figure \ref{values}), in which the H$_{2}$ has not been considered. Nevertheless, we checked that taking the simplified model does not affect our global modeling, i.e., the values of spectral slopes and extinction.}, while the value in the host galaxy line-of-sight $N_{\rm H,X}$ is left as a free parameter of the model. Following the discussion of \citet{Watson2013} we assume solar abundances of \citet{Anders1989}.

In principle, we do not know whether the spectral break is sharp as in Equation \ref{eq:powerlaw} or the transition is more mild and smooth \citep[e.g.,][]{Granot2002,Uhm2014}. Indeed, at least in one case the cooling break appears to be smooth: \citet[][]{Filgas2011} found that multiepoch SEDs of GRB\,091127 can be described with a broken power-law with a smooth spectral transition. We carefully evaluated the effect of the smoothness on our modeling of broadband SEDs with data of X-shooter quality. We find that the available data cannot constrain the magnitude of the smoothness (i.e., the sharpness index in Equation 1 in \citealt{Filgas2011}). Furthermore, if a break is very smooth and lies near the optical region, the values of other parameters (like $A_{\rm V}$ and $\beta_{\rm O}$) obtained in the modeling are considerably affected and uncertain. This is not the case if the break is sharp. In the absence of a better knowledge on the spectral smoothness, we assume that the breaks are sharp in order to reduce the possible systematic parameter errors.

For GRBs occurring at $z \gtrsim 2$, the host's Ly$\alpha$ absorption line enters the X-shooter observational window. In order to better constrain the UV slope of the extinction curve we decided to include the modeling of the red wing of the Ly$\alpha$ line in our analysis. The Ly$\alpha$ line is characterized by its central wavelength $\lambda_{\rm Ly\alpha}$, column density of the absorbing gas $N_{\rm HI}$ and Doppler parameter $b$. As expected for the damped Lyman alpha absorbers,  the fitting turned out to be very insensitive to the latter, therefore we fixed its value to $b = 12.6$ km s$^{-1}$, which is the average Doppler parameter of GRB host galaxy absorption lines \citep{Postigo2012}. To model the Ly$\alpha$ line we used the analytical approximation derived by \citet{Garcia2006,Garcia2007}. The final model applied to the data and already corrected for Galactic extinction and photoelectric absorption can therefore be summarized as:

\begin{equation}
F_{\rm \nu,obs} = F_{\nu}\times10^{-0.4A_{\lambda} }\times\exp[-N_{\rm H,X}\sigma(\nu)]\times\exp[-\tau_{\rm Ly\alpha}(\nu)],
\end{equation}

where $\sigma(\nu)$ is a cross-section for photoelectric absorption occurring from the gas in the host galaxy.

The SED fitting was carried out with the spectral fitting package XSPECv12.8 \citep{Arnaud1996}.The analysis was done separately for $(i)$ the X-shooter spectrum and $(ii)$ the broadband SED including the X-shooter and X-ray spectra. We also modelled a broadband SED with photometric data instead of the X-shooter spectrum, in order to compare the two types of analyses. Confidence intervals were computed at 90$\%$ confidence level following \citet{Avni1976} and \citet{Cash1976} with one parameter of interest. Confidence intervals were being computed independently for each model parameter. We did not investigate possible correlations of uncertainties between the model parameters. We considered a fit to be successful if the null probability (that is, the probability of getting a value of $\chi^2$ as large or larger than observed if the model is correct) was better than 10$\%$. If the broken power-law provided a better fit than a single power-law, we used the $F$-test (with the probability of $5\%$ as a threshold) to assess whether the improvement in the $\chi^2$ is statistically significant. 

\section{Results}
\label{results}
   
Best-fitted model parameters for the SEDs of the GRBs in our sample are reported in Table \ref{bestfit}. An example of the best model in the case of GRB\,120815A is plotted in Figure \ref{sed1}; plots corresponding to other GRBs are shown in Figures \ref{figsed1} and \ref{figsed2}. Detailed fitting results are outlined in Table \ref{sedresults} and discussed in this Section on a case by case basis. 

In the following we report qualitative fitting information for each analysed SED in the sample. We also briefly discuss the compatibility of our SED results with the closure relations for each case, assuming models assembled by \citet[][Table 1]{Racusin2009}. To simplify the discussion we do not consider structured jet models and models with continuous energy injection. We caution that for a detailed understanding of each event, the reader should consult works dedicated to each GRB. 

\subsection{GRB\,100219A}

Photometric data used for normalization and photometric SED analysis were obtained from \citet{Thoene2013}. The X-shooter SED alone can be fitted by a single power-law and SMC or LMC extinction curve. The fit is improved when we include X-ray data - the broadband SED can be fitted by a power-law and all three extinction curves. The best fit is achieved with the LMC extinction curve. A broken power-law does not statistically improve the fit, therefore we do not consider it further. This is a high-redshift GRB ($z = 4.667$), therefore we had to include the Ly$\alpha$ absorption to constrain the UV slope. The value of the $\log N_{\rm HI} = 21.0 \pm 0.1$ that we obtain in the fit is comparable within errors with the one derived by \citet{Thoene2013} from normalized spectrum (i.e., $\log N_{\rm HI} = 21.14 \pm 0.15$). Fixing the value to $\log N_{\rm HI} = 21.14$ results in a bit worse $\chi^2$ statistics, but otherwise the other parameter values do not change.

For the photometric SED we use only filters not affected by host's Ly$\alpha$ and Lyman forest. According to the statistics, broadband SED with photometric optical points is best-fitted with a power-law and MW curve. However, visual inspection shows that the optical SED part is better described with a power-law and SMC or LMC extinction. The latter two give a comparable power-law slope and slightly lower extinction $A_{\rm V}$ as the one obtained with the X-shooter broadband SED. A rather low reduced $\chi^2$ values are due to small number of data points (both at optical and X-rays) included in the fit.

Optical light curve is characterized by several moderate rebrightenings \citep{Mao2012}. The X-shooter spectrum is taken right after a bump peaking at $\sim$ 20 ks. The light curve slope at the X-shooter epoch is $\alpha \approx 1.31$ \citep{Thoene2013}. The case of an ISM environment and $\nu_{\rm c} > \nu_{\rm X}$ seems to describe the optical light curve the best ($\alpha \sim 1.1$, $p \sim 2.5$). At this point the X-ray light curve is in a transition from a shallow to steep ($\alpha \sim 2.9$) phase, which is not seen in the optical \citep{Thoene2013}. If this is due to a geometrical effect (i.e., a jet break) rather than spectral evolution, our result is wrong. Due to a combination of aforementioned bump and sparse data it is hard to estimate the correction to normalization of the X-ray spectrum due to a jet break. We approximately estimate that X-ray afterglow at the X-shooter epoch would be a factor of 2 brighter in the absence of a jet break. After applying this correction and repeating the fit, we find that a single power-law and the LMC-type extinction curve still describe the data the best, though the extinction ($A_{\rm V} \approx 0.26$) and spectral slope ($\beta \approx 0.66$) do change a bit, as expected.

\begin{table*}[t]
\renewcommand{\arraystretch}{1.5}
\caption{Results of the best-fit models of the broadband SEDs}
\centering
\begin{tabular}{lcccccccr}
\hline\hline
GRB & Extinction & $N_{\rm H,X}$                       & $\beta_{1}$ & $\beta_{2}$ & $\nu_{\rm break}^{(a)}$ & $\log N_{\rm HI}$(Ly$\alpha$) & Av &  $\chi^{2}$/dof\\
       &    type       &  $[10^{22} {\rm cm}^{-2}]$ &                    &                    & $[10^{15}$ Hz]               &   $\log$ cm$^{-2}$                       &                          &\\
\hline
100219A & LMC & < 6.2 & 0.73$_{-0.02}^{+0.02}$ & & & 21.0$_{-0.1}^{+0.1}$ & 0.23$_{-0.02}^{+0.02}$ & 34.8/30 \\
100418A & SMC &  0.14$_{-0.12}^{+0.21}$ & 0.73$_{-0.08}^{+0.07}$ & 1.04$_{-0.03}^{+0.03}$ & 0.6$_{-0.1}^{+0.3}$ & & 0.20$_{-0.02}^{+0.03}$ & 20.8/23\\
100814A & SMC &  0.35$_{-0.11}^{+0.13}$ & 0.52$_{-0.07}^{+0.07}$ & 1.05$_{-0.02}^{+0.02}$ & 2.3$_{-0.1}^{+3.8}$ & & 0.20$_{-0.03}^{+0.03}$ & 70.8/66\\
100901A & SMC & 0.25$_{-0.18}^{+0.23}$ & 0.50$_{-0.04}^{+0.04}$ & 1.06$_{-0.06}^{+0.05}$ & 5.8$_{-3.2}^{+8.8}$ & & 0.29$_{-0.03}^{+0.03}$ & 44.2/41\\
120119A & LMC & 1.98$_{-0.40}^{+0.50}$ & 0.89$_{-0.01}^{+0.01}$ & & & 23.4$_{-0.2}^{+0.2}$ & 1.07$_{-0.03}^{+0.03}$ & 106.0/81\\
120815A & SMC & 0.66$_{-0.39}^{+0.52}$ & 0.38$_{-0.05}^{+0.07}$ & 0.84$_{-0.02}^{+0.02}$ & 1.4$_{-0.8}^{+0.7}$ & 22.3$_{-0.2}^{+0.2}$ & 0.32$_{-0.02}^{+0.02}$ & 26.0/47\\
130427A & SMC  & 0.08$_{-0.02}^{+0.02}$ & 0.37$_{-0.04}^{+0.05}$ & 0.68$_{-0.01}^{+0.01}$ & 0.7$_{-0.2}^{+0.3}$ & & 0.16$_{-0.02}^{+0.02}$ & 129.3/147\\
130603B & SMC  & 0.20$_{-0.09}^{+0.15}$ & 0.42$_{-0.22}^{+0.12}$ & 0.92$_{-0.04}^{+0.08}$ & 0.8$_{-0.1}^{+0.1}$ & & 1.19$_{-0.12}^{+0.23}$ & 21.3/23\\
130606A$^{(b)}$ & / & < 3.5 & 0.96$_{-0.02}^{+0.02}$& & & 19.9$_{-0.2}^{+0.2}$ & < 0.01 & 48.8/31\\ 
\hline
\end{tabular}
\tablefoot{Detailed summary of the fitting results is outlined in Table \ref{sedresults} and discussed in Section \ref{results}. \newline (a) Host rest-frame value. \newline (b) GRB\,130606A is found to be consistent with $A_{\rm V} \sim 0$, therefore no extinction curve is needed for modeling.}
\label{bestfit}
\end{table*}

\subsection{GRB\,100418A}

Photometric data used for normalization and photometric SED analysis were obtained from de Ugarte Postigo et al. (in prep). Three afterglow spectra were obtained at $\sim 0.4, 1.4$ and 2.4 days after the burst. Due to the bad flux calibration of the first epoch spectra and possible contamination from host galaxy and supernova emission of the third epoch of observation we only use the second epoch spectrum. The X-shooter SED alone can be fitted by a single power-law. However, the broadband SED requires a broken power-law shape. The difference in spectral slopes is $\Delta \beta = 0.31 \pm 0.09$. Fixing the spectral difference to $\Delta\beta = 0.5$ significantly changes only $\beta_{\rm O}$, while other parameter values stay almost unchanged - the $F$-test probability ($< 1\%$) confirms that the model in which both spectral slopes are left free to vary is statistically better. While the $\chi^2$ statistics is similar for all three extinction types, visual inspection reveals that the SMC extinction curve provides the best fit to the blue part of the optical data. 

The broadband SED with photometric optical points is fitted equally well with all three extinction curves and a broken power-law. The best-fit slopes are  comparable to the values obtained from the X-shooter broadband SED fit, while extinction values are slightly higher (but within errors). The broadband fit with a single power-law also describes the data well, however the F-test probability ($< 5\%$) confirms the broken power-law improves the fit significantly.

Initially shallow evolution of the optical light curve was followed by a rebrightening, reaching its peak brightness at $\sim 0.6$ days \citep{Marshall2011}. The data from de Ugarte Postigo et al. (in prep) suggest an optical late-time steepening of $\alpha_{\rm O} \sim 1.5$, similar to the X-ray decay of $\alpha_{\rm X} \sim 1.4$ in this late phase. \citet{Marshall2011} find a shallower optical steepening of $\alpha_{\rm O} \sim 1.1$. In the latter case the difference between the optical and X-ray light curve slopes cannot be explained within the standard model without a spectral break between the optical and X-ray region. On the other hand, the case of $\alpha_{\rm O} \sim 1.5$ could be explained within a model of non-spreading uniform jet in a wind environment. We do not find any significant color evolution in a few days around the X-shooter epoch (de Ugarte Postigo et al. in prep), which is in contradiction with the stellar wind environment scenario.

\subsection{GRB\,100814A}

Photometric data used for normalization and photometric SED analysis were obtained from \citet{Nardini2014}. Three afterglow spectra were obtained at $\sim 0.038, 0.089$ and 4.1 days after the burst. Due to the discrepancy between flux calibrated spectra and photometric measurements at the first two epochs, only the spectrum taken 4.1 days after the burst has been analysed. The X-shooter SED alone can be fitted with a SMC extinction curve and a power-law intrinsic behavior. Statistically there is no need for a spectral break. LMC and MW curves fail to reproduce the data. The broadband SED is fitted well with a broken power-law and the SMC or LMC extinction curve. However, the LMC clearly overpredicts the 2175 $\mathrm{\AA}$ bump. We therefore prefer the SMC extinction curve as the best model to describe the data. The difference in spectral slopes is $\Delta \beta = 0.53 \pm 0.08$. Fixing the spectral difference to $\Delta\beta = 0.5$ does not change the results significantly.

The broadband SED with photometric optical points is fitted equally well with all three extinction curves and a broken power-law. The position of the spectral break is similar to the position obtained with the X-shooter broadband SED, while the spectral slopes are different. The broadband fit with a single power-law is bad ($\chi^{2}/{\rm dof} > 3.0$).

Optical afterglow light curve of GRB\,100814A is characterized by a strong rebrightening at $\sim 20$ ks. At the X-shooter epoch the optical and X-ray light curves decay as $\alpha \approx 2.25, 2.30$, respectively \citep{Nardini2014}. Our SED results are marginally consistent with the model of a spreading uniform jet. Light curve evolution in optical prior to the X-shooter epoch is chromatic - afterglow is becoming redder with time \citep{Nardini2014}. If the position of the cooling frequency is indeed between the optical and X-ray (as we find it to be), then this would suggest that the frequency is traveling towards higher frequencies.

\subsection{GRB\,100901A}

Photometric data used for normalization  and photometric SED analysis were obtained from Gomboc et al. (in prep). The X-shooter SED alone can be fitted by a single power-law spectrum. Broadband fit reveals the need for a spectral break between the optical and X-ray regions. SMC extinction curve provides an excellent fit to the data, while the LMC and MW curves are completely inadequate to describe the SED. The difference in spectral slopes is $\Delta \beta = 0.56 \pm 0.07$. Fixing the spectral difference to $\Delta\beta = 0.5$ does not change the results significantly.

The broadband SED with photometric optical points is fitted best with a broken power-law. Model with the SMC-type dust provides a marginally better fit than LMC- or MW-type dust. The values differ quite a lot from the ones obtained from X-shooter broadband SED, owing to the unavailable near infrared photometric observations. 

The afterglow exhibits an extreme rebrightening phase at optical wavelengths (e.g., \citealt{Hartoog2013}, Gomboc et al. in prep.). Light curve at the time of the X-shooter observation evolves practically achromatically with $\alpha_{\rm O} = 1.52 \pm 0.05$ and $\alpha_{\rm X} =  1.55 \pm 0.05$. None of the closure relations can simultaneously describe the observed spectral and temporal properties at late times.

\begin{figure*}[htp]
\centering
\begin{tabular}{cc}
\includegraphics[scale=0.38]{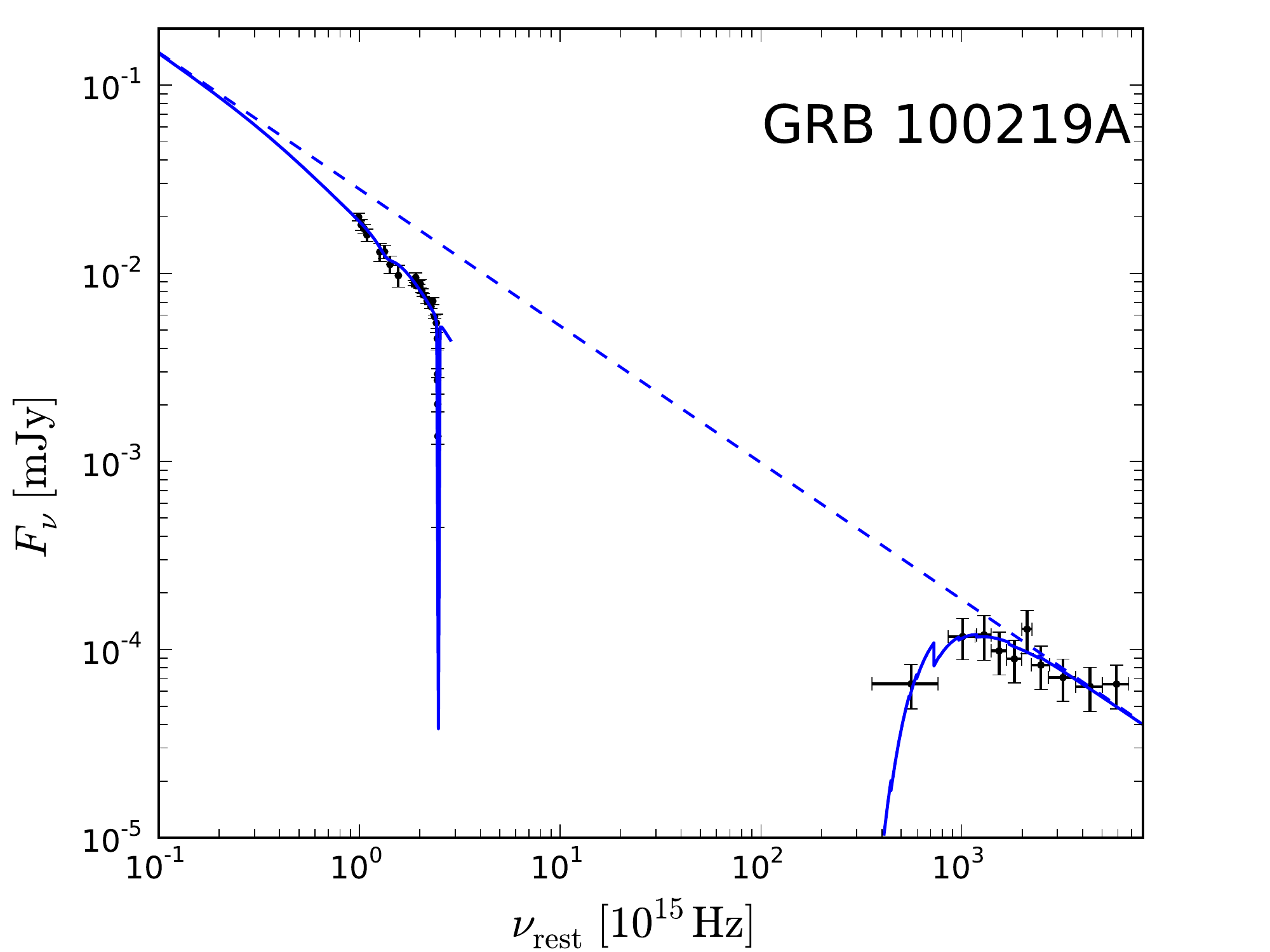}&
\includegraphics[scale=0.38]{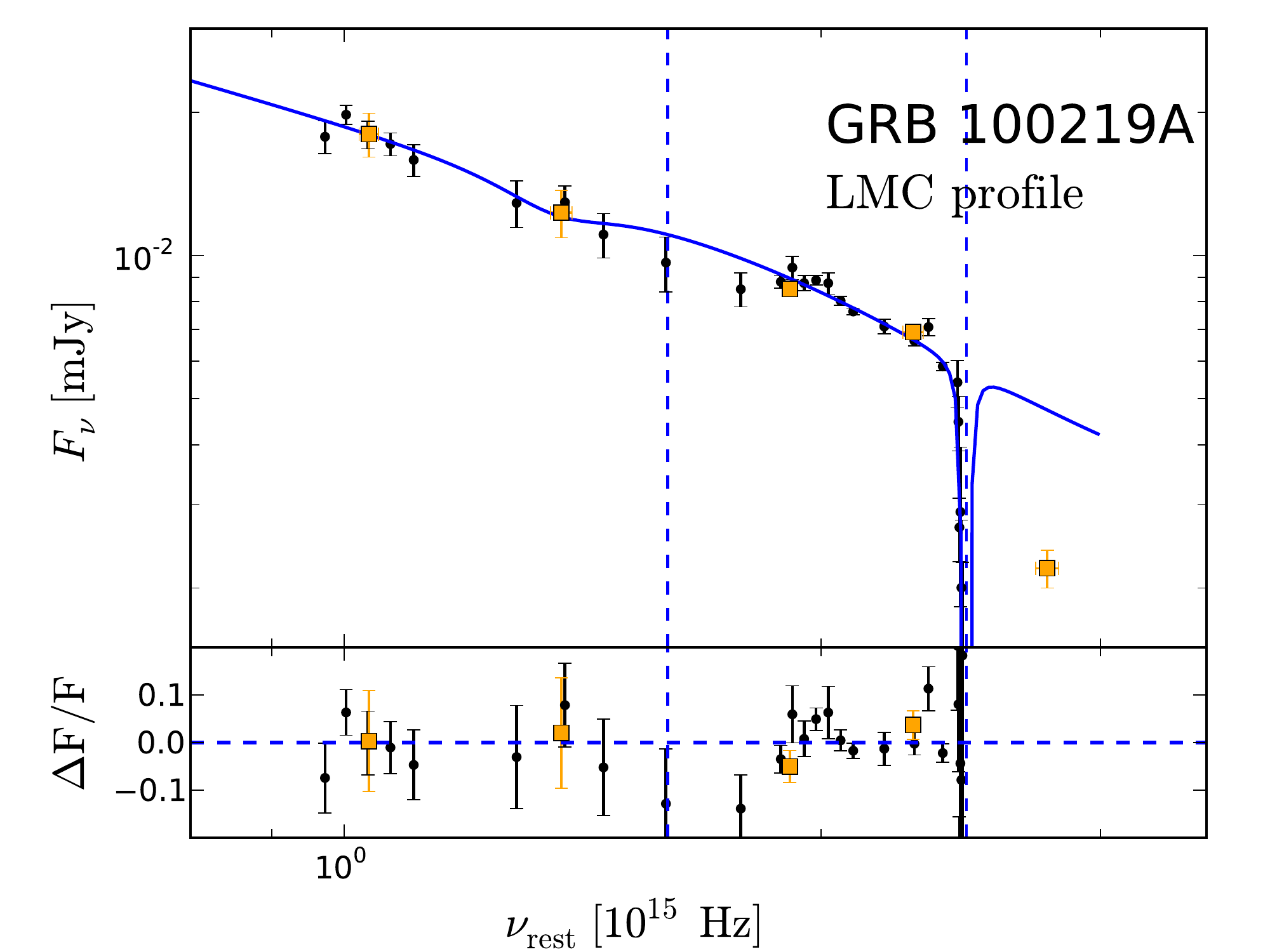}\\
\includegraphics[scale=0.38]{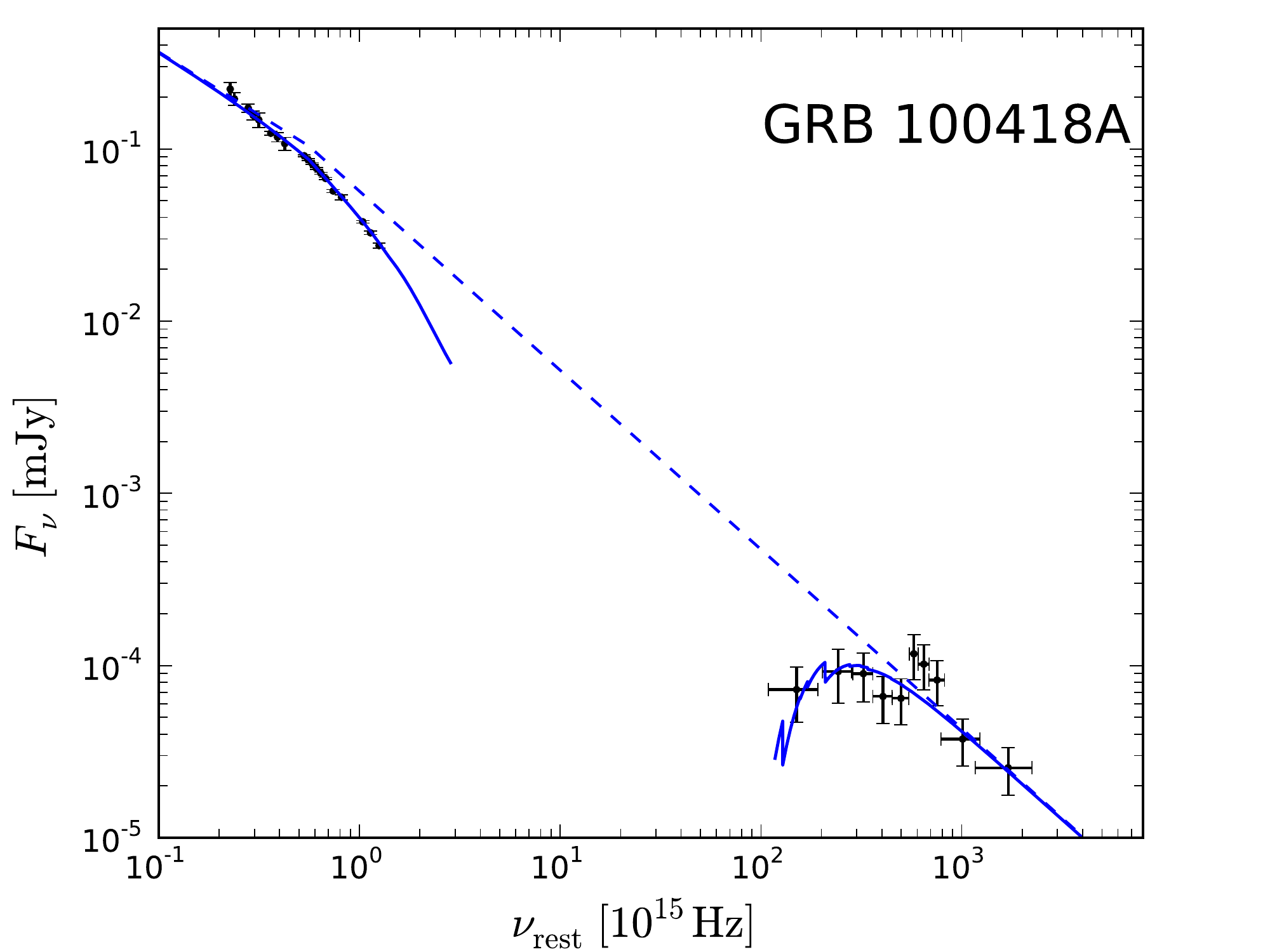}&
\includegraphics[scale=0.38]{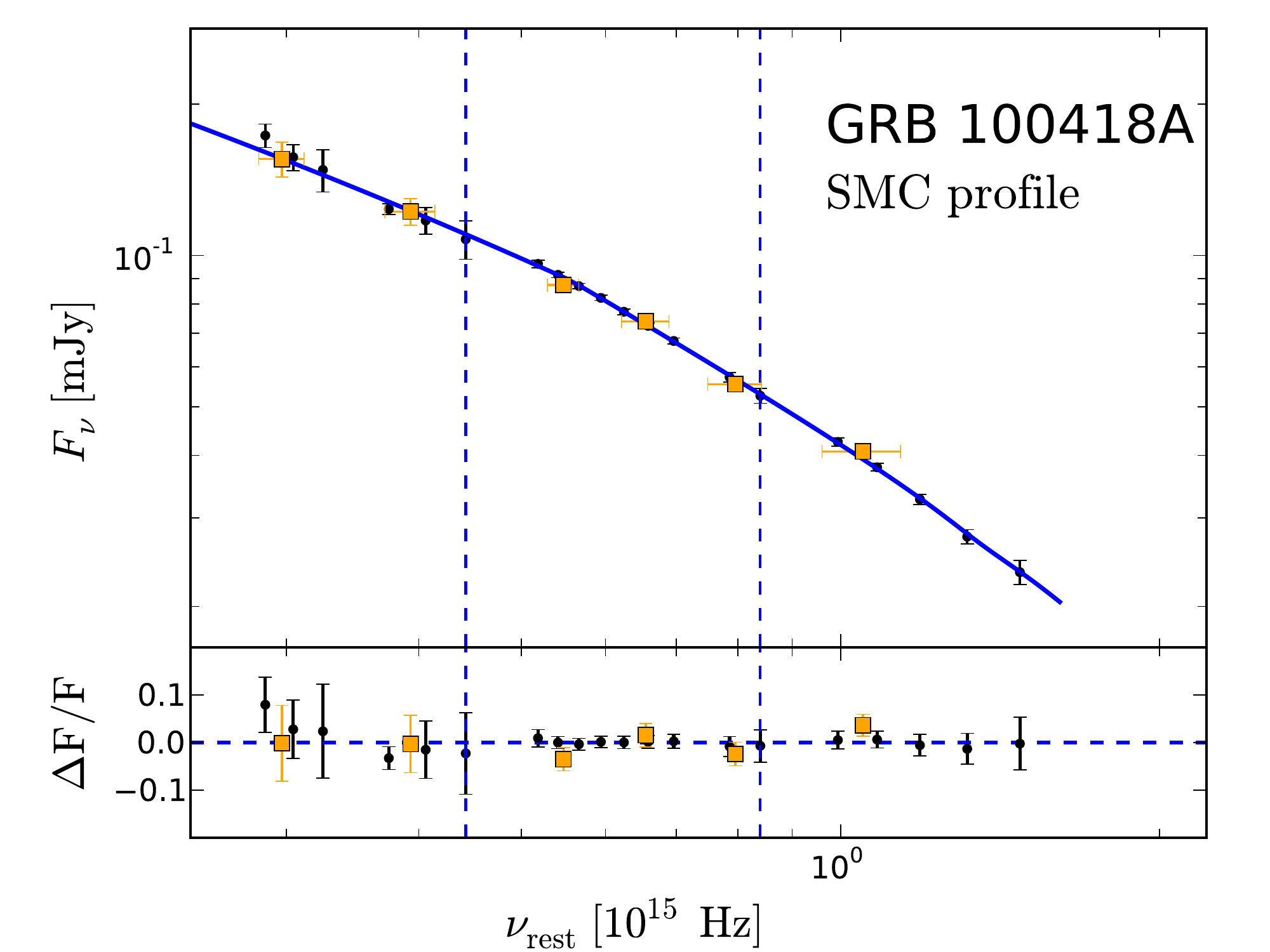}\\
\includegraphics[scale=0.38]{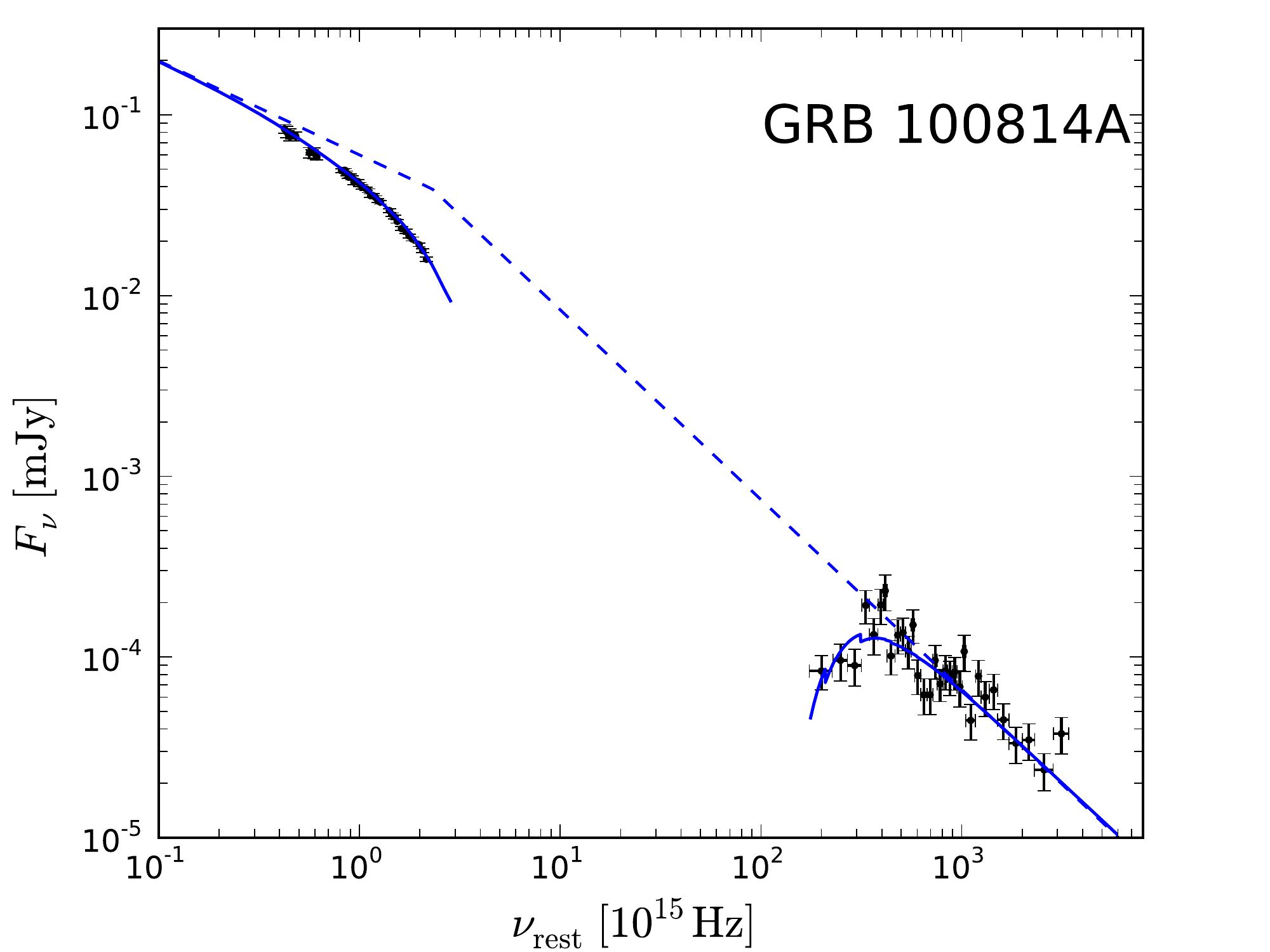}&
\includegraphics[scale=0.38]{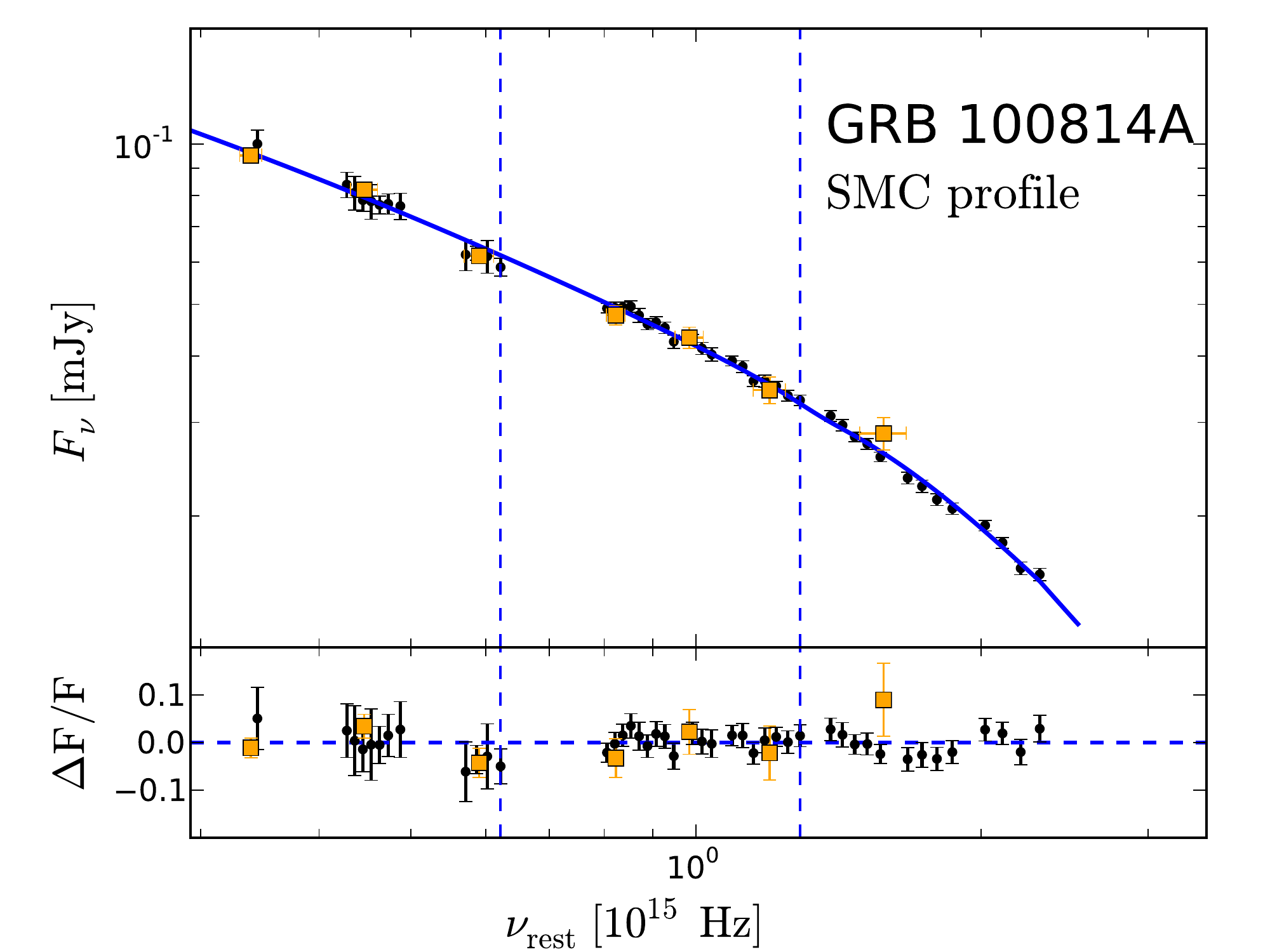}\\
\includegraphics[scale=0.38]{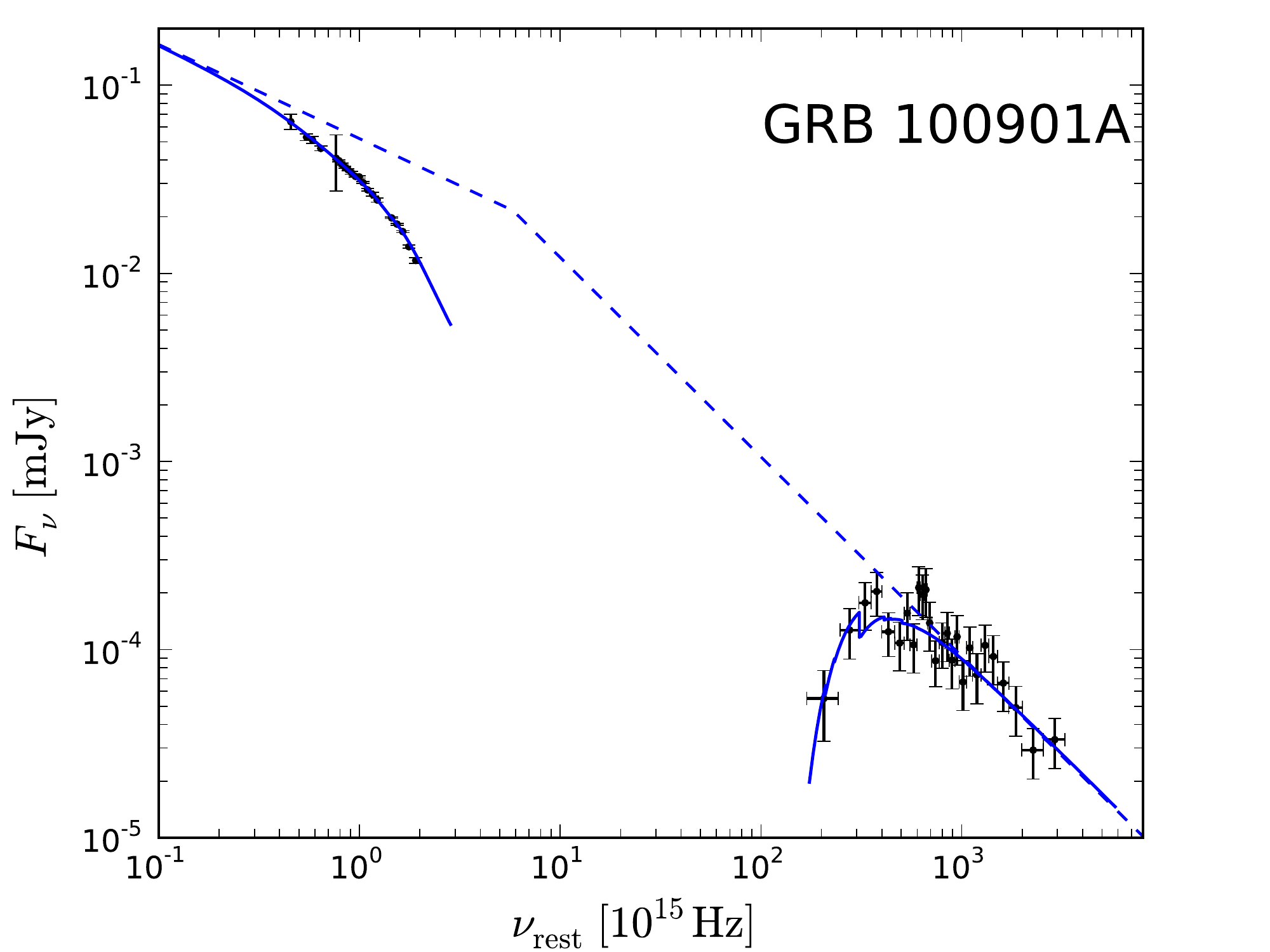}&
\includegraphics[scale=0.38]{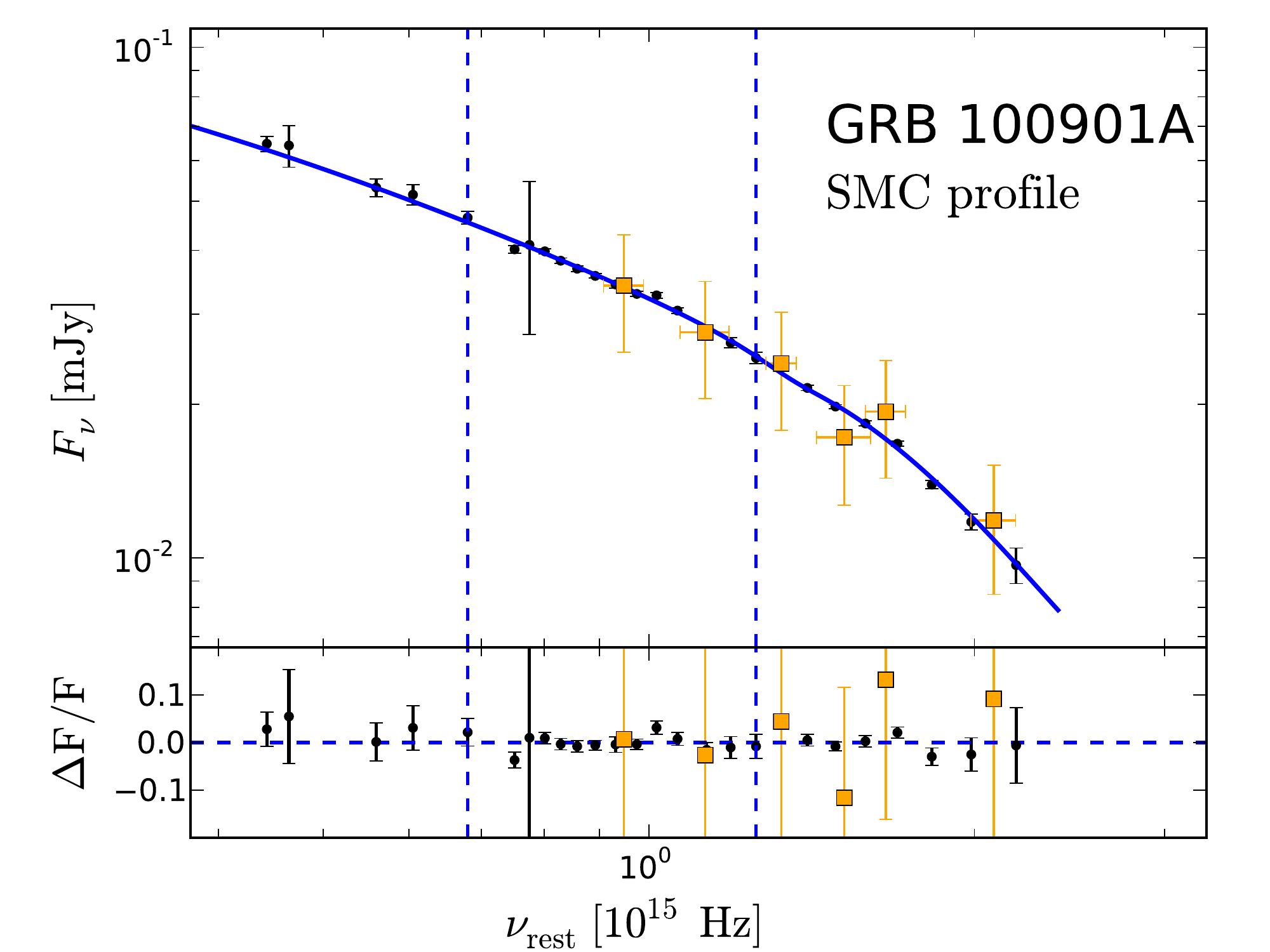}\\
 \end{tabular}
\caption{SEDs and best-fitted models of the sample. For details, see caption in Figure \ref{sed1}.}
 \label{figsed1}
\end{figure*}   

\begin{figure*}[htp]
\centering
\begin{tabular}{cc}
\includegraphics[scale=0.38]{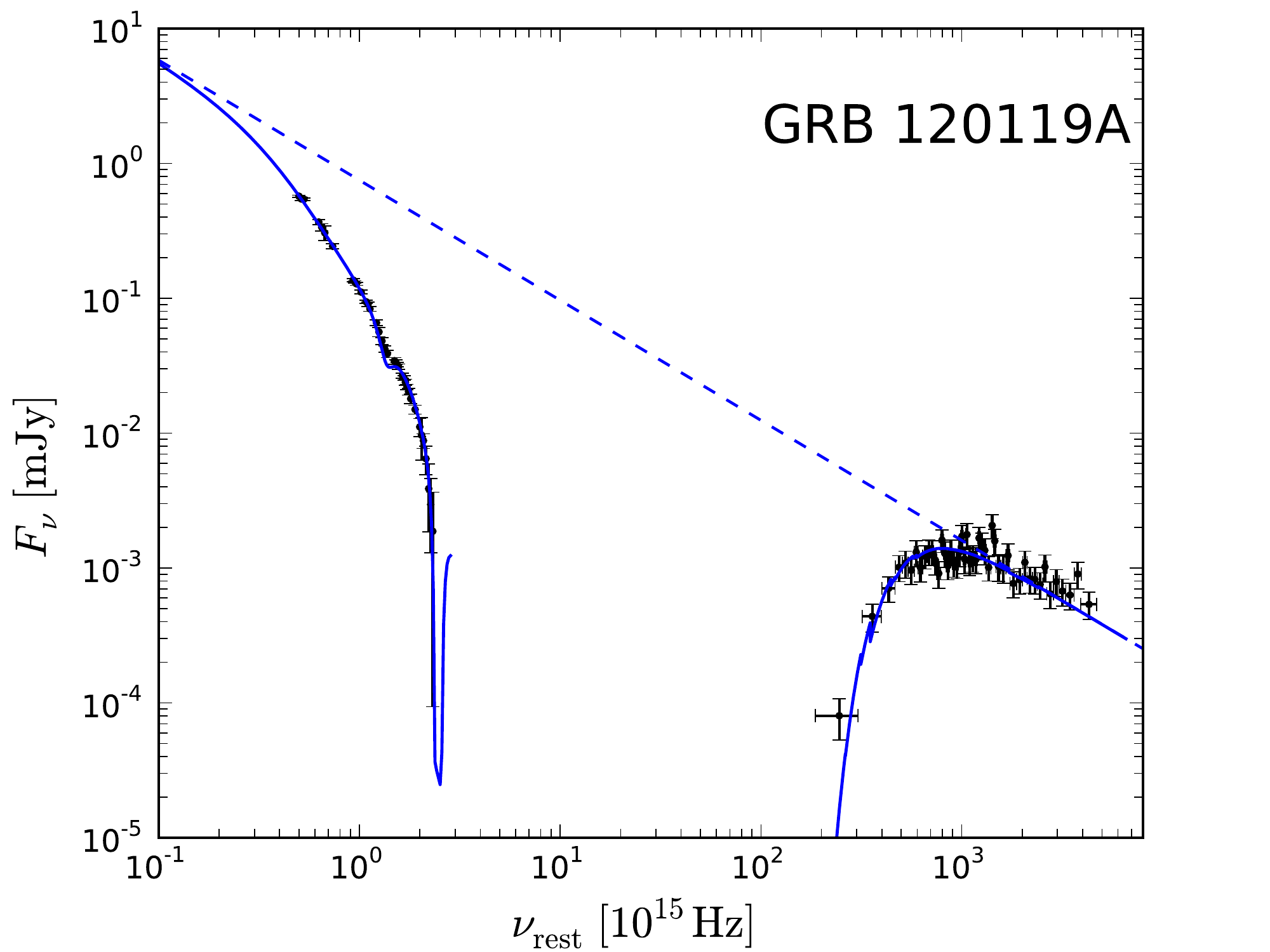}&
\includegraphics[scale=0.38]{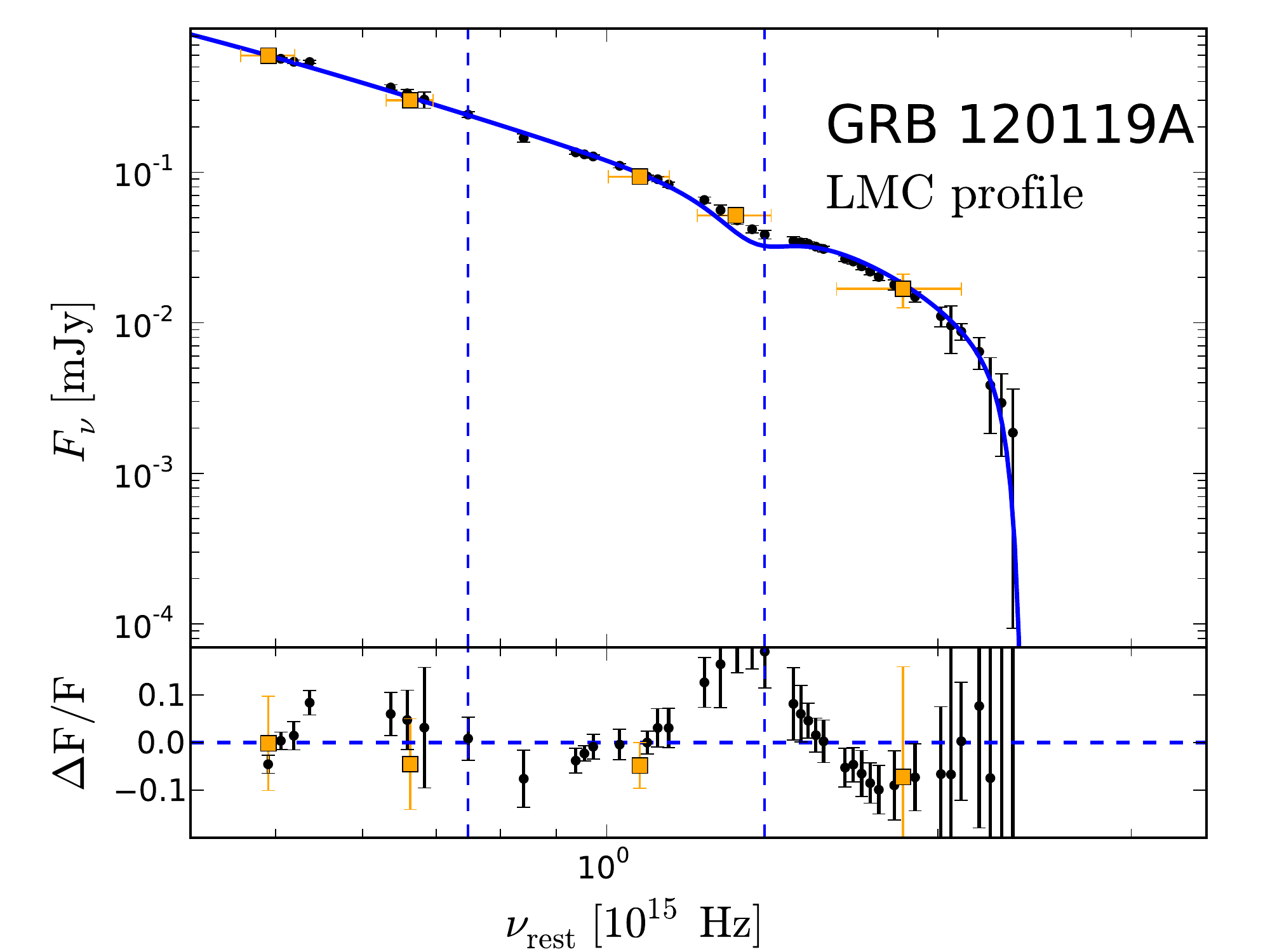}\\
\includegraphics[scale=0.38]{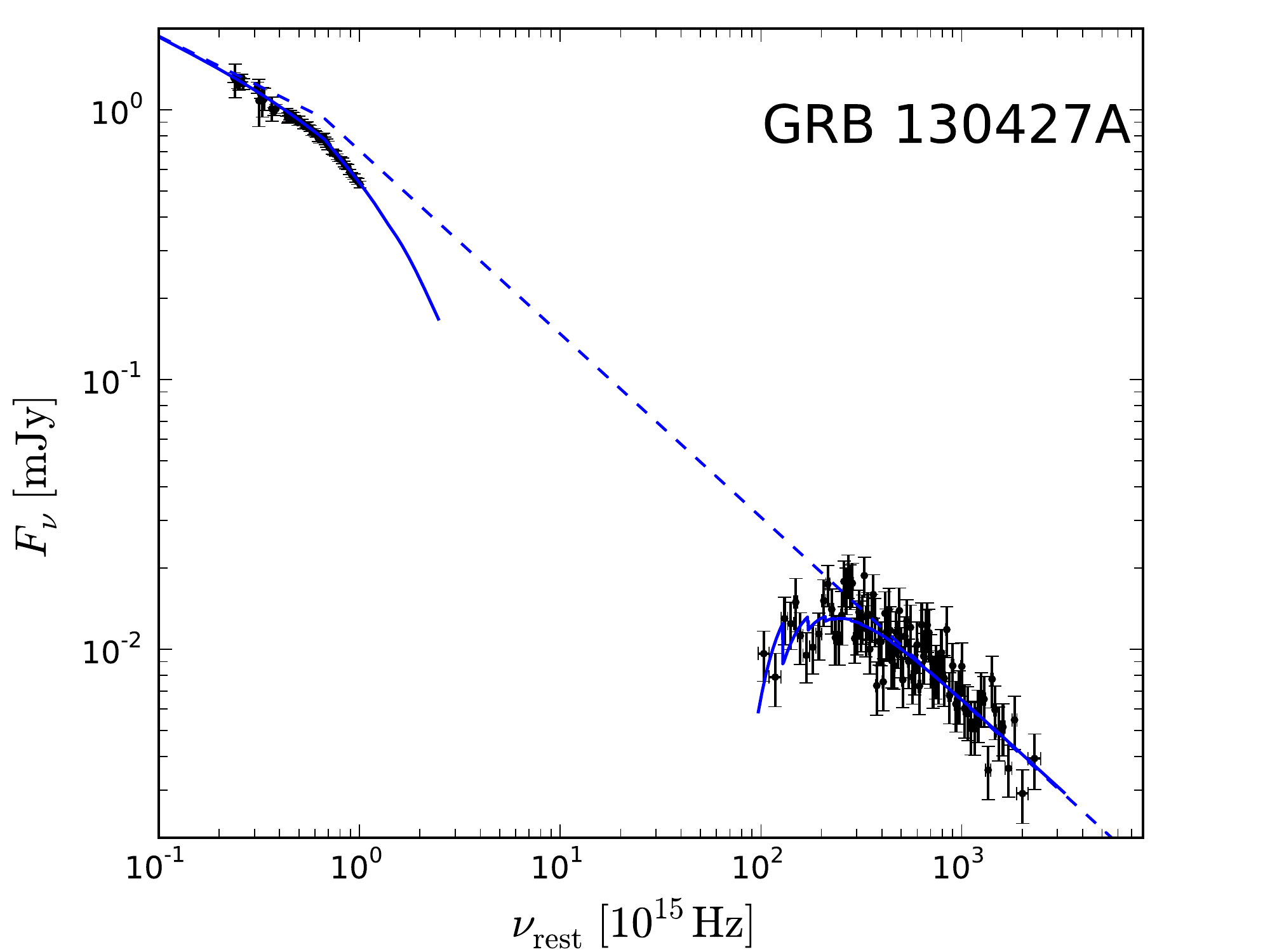}&
\includegraphics[scale=0.38]{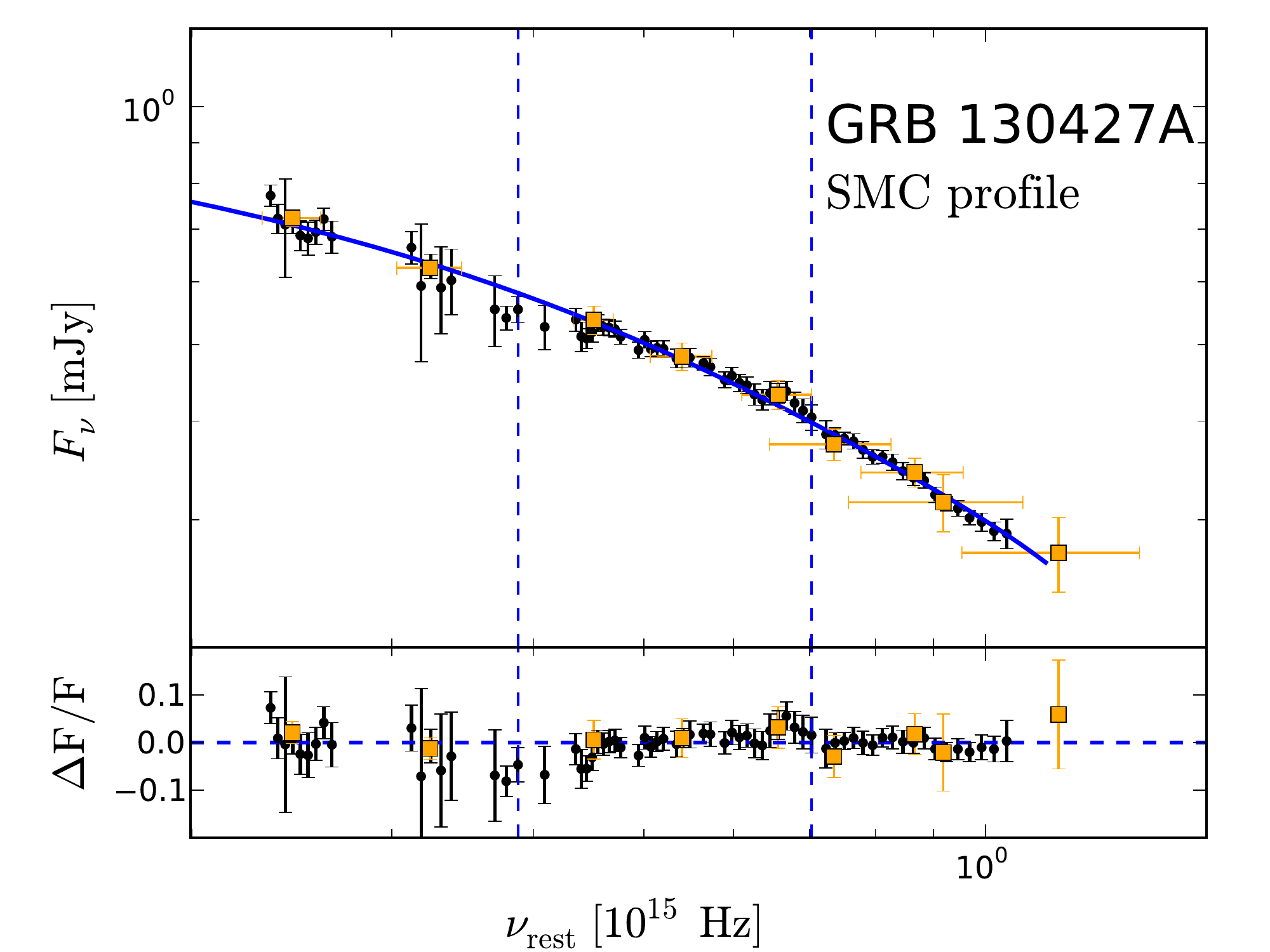}\\
\includegraphics[scale=0.38]{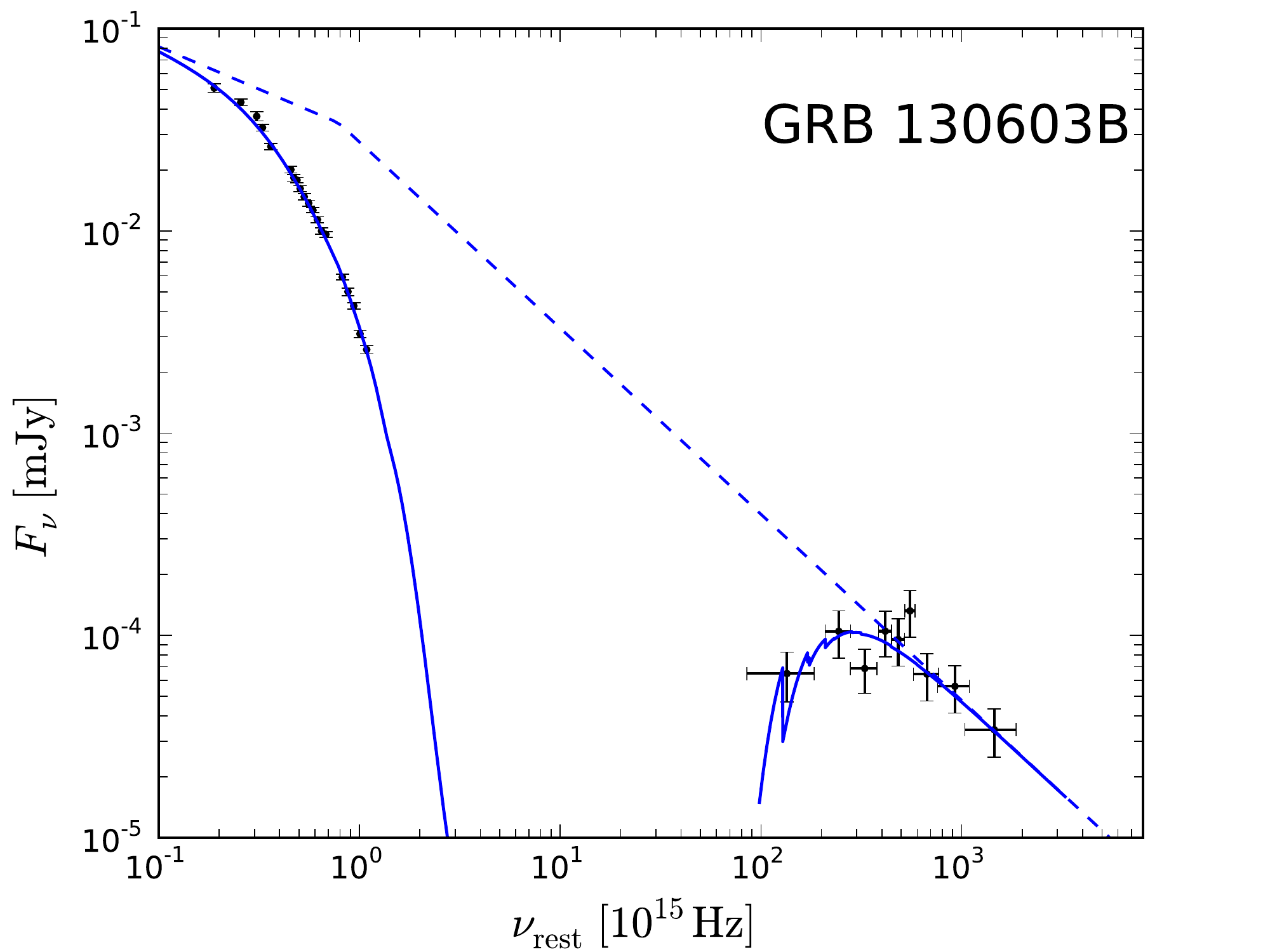}&
\includegraphics[scale=0.38]{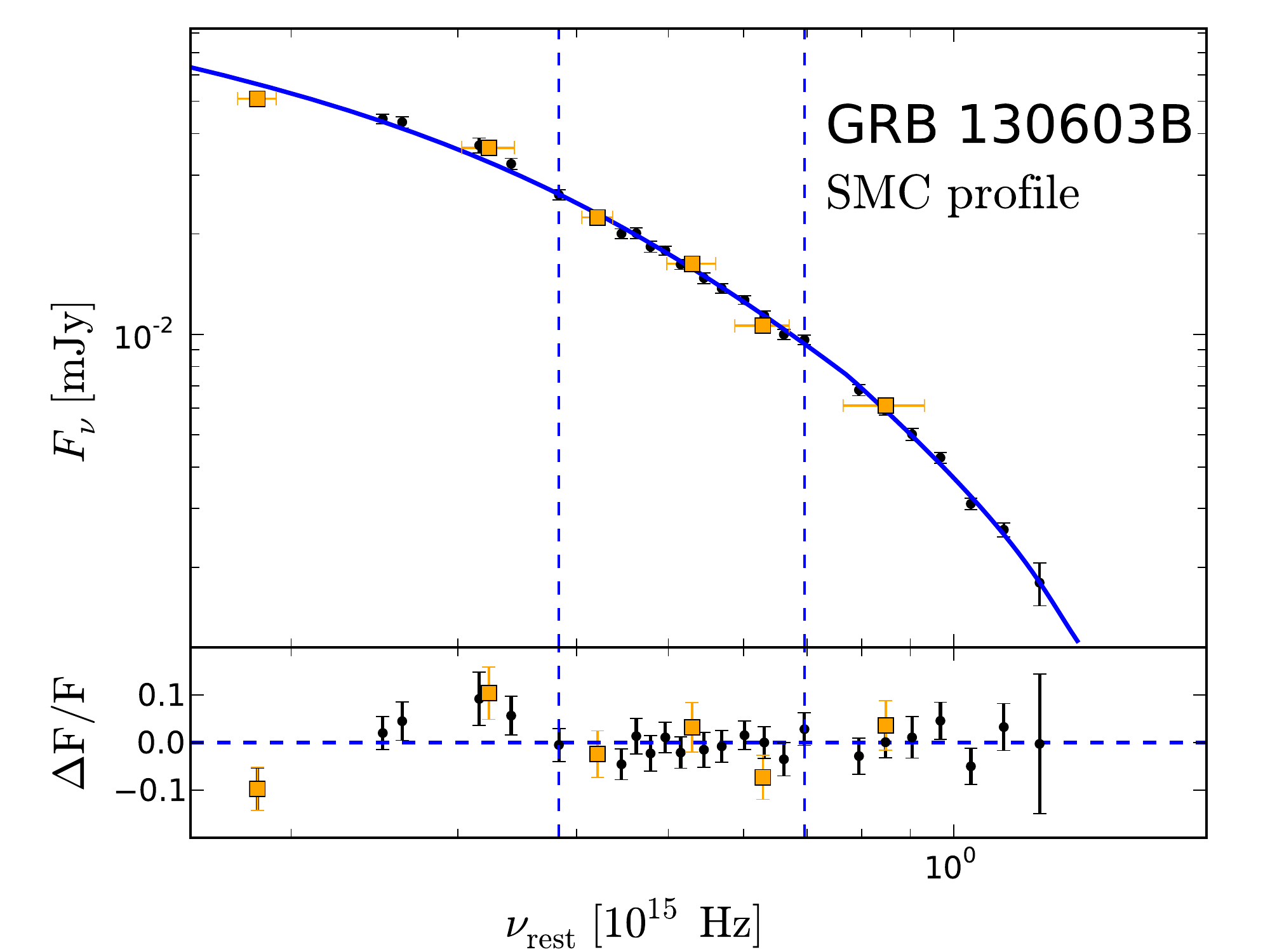}\\
\includegraphics[scale=0.38]{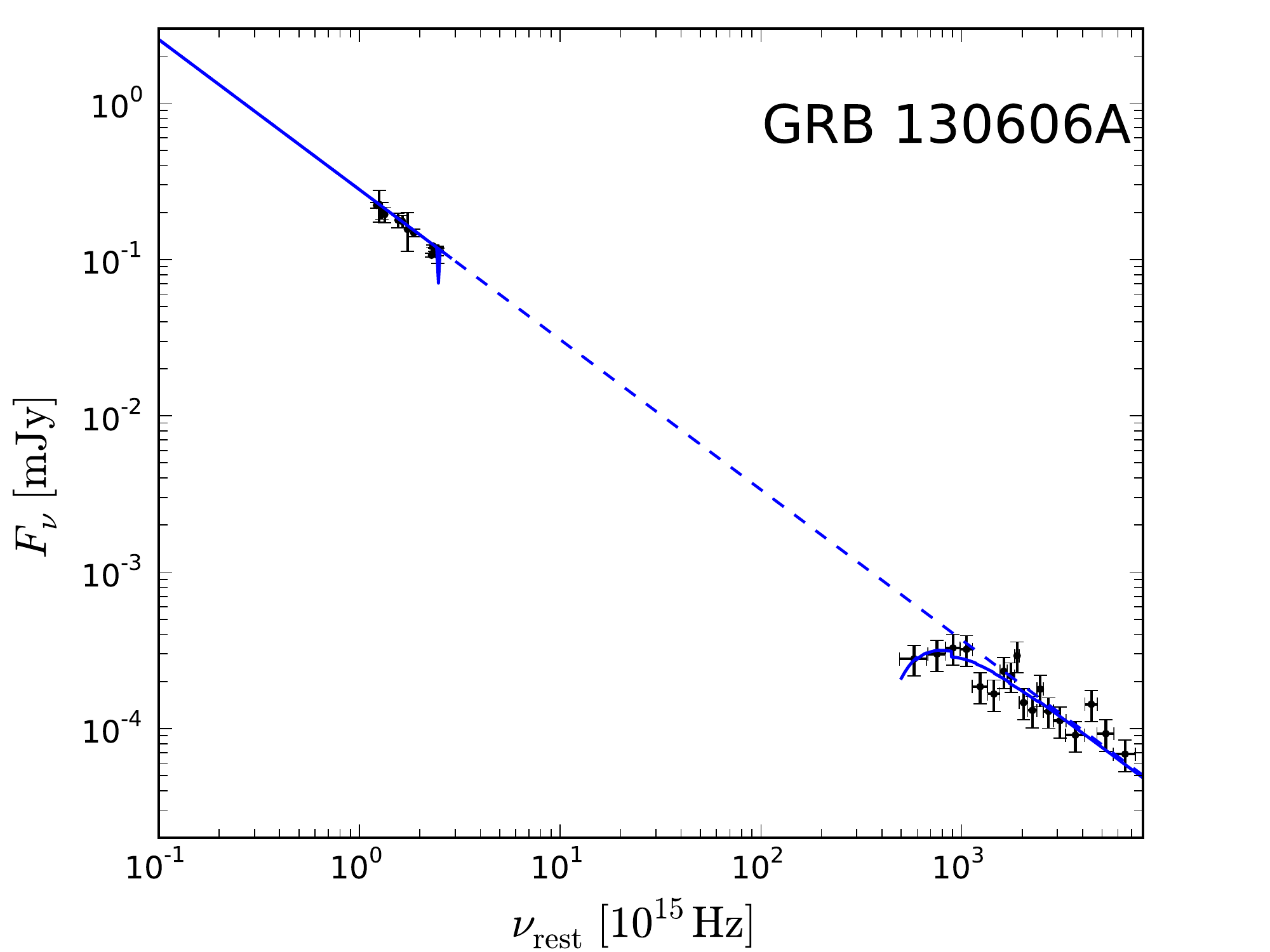}&
\includegraphics[scale=0.38]{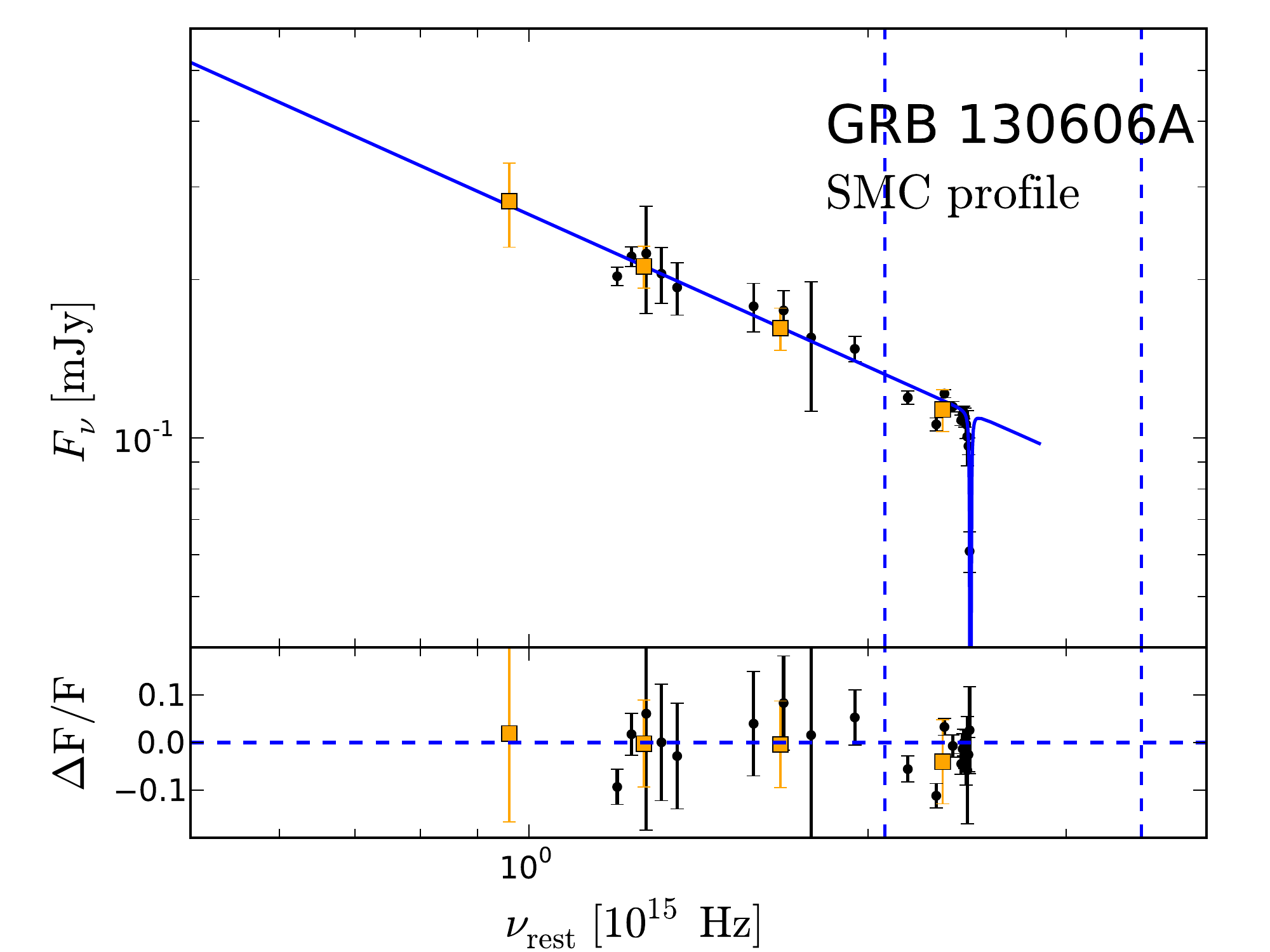}\\
 \end{tabular}
\caption{SEDs and best-fitted models of the sample. For details, see caption in Figure \ref{sed1}.}
 \label{figsed2}
\end{figure*}   

\subsection{GRB\,120119A}

Photometric data for normalization and photometric SED analysis were obtained from \citet{Morgan2014}. Two afterglow spectra were obtained at $\sim 0.074$ and $0.20$ days after the burst. Due to poor signal-to-noise of the second epoch spectrum we only use the one obtained at 0.074 days. This is the only case in our sample with a clear 2175 $\mathrm{\AA}$ bump in the SED. In addition, the blue part of the SED is already affected by the red wing of the Ly$\alpha$ line. We cannot obtain a very good fit modeling the X-shooter SED alone. The broadband SED is fitted best with a power-law and LMC extinction curve. Even so, the LMC curve overpredicts the strength of the 2175 $\mathrm{\AA}$ bump. A broken power-law on a broadband dataset (and X-shooter data alone) results in an unphysical result of $\beta_{2} < \beta_{1}$. Ly$\alpha$ line lies in the bluest part of the spectra, where signal-to-noise is very low. This is probably the main factor contributing to the overestimated value of $\log N_{\rm HI} = 23.4 \pm 0.2$ with respect to the value measured from the normalized spectra (e.g., $\log N_{\rm HI} = 22.5 \pm 0.3$; Vreeswijk et al. in prep.).

The broadband SED with photometric optical points is fitted best with a single power-law and LMC extinction curve. Best-fit spectral slope is similar to the one obtained with the X-shooter broadband fit, while the extinction $A_{\rm V}$ is slightly lower. Statistically there is no need for a spectral break in the fitted spectral region.

The early time light curve is characterized by color evolution, possibly explained as a result of dust destruction \citep{Morgan2014}. At the X-shooter epoch the light curve evolution is achromatic in the optical-to-X-ray spectral range with $\alpha \sim 1.3$. Spectral and temporal indices are consistent with the case of $\nu_{\rm C} > \nu_{\rm X}$ and an ISM environment.

\subsection{GRB\,120815A}

Photometric data used for normalization and photometric SED analysis were obtained from \citet{Kruhler2013}. The X-shooter SED can be fitted with a broken power-law and SMC extinction curve. Models with LMC and MW extinction curves are successful in reproducing the data, but the values of the post-break slopes are unrealistically steep (given the X-ray part of the SED). The broadband SED can only be fitted with a broken power-law and an SMC extinction curve. This GRB originates at $z = 2.358$, therefore we had to include the Ly$\alpha$ absorption to constrain the UV slope. The value of the $\log N_{\rm HI} = 22.3 \pm 0.2$ that we obtain is slightly larger than the one derived by \citet{Kruhler2013}, who found $\log N_{\rm HI} = 21.95 \pm 0.10$. As discussed by \citet{Kruhler2013}, the Ly$\alpha$ line is contaminated by vibrationally excited H$_{2}^{*}$ lines that form a continuum around this spectral region and cause the line to appear stronger. Not taking the H$_{2}^{*}$ into account, the column density value of \textit(contaminated) Ly$\alpha$ is $\log N_{\rm HI} \approx 22.1$, closer to our value. Fixing the value to $\log N_{\rm HI} = 22.1$ in the modeling does not change the results (see Table \ref{sedresults}). The difference in spectral slopes is $\Delta \beta = 0.46 \pm 0.07$: fixing it to $\Delta \beta = 0.5$ does not change the results of the fitting significantly.

We did not use the $g'$-band photometric point in the photometric SED analysis because the measurement is affected by Ly$\alpha$ absorption. Broadband SED with photometric optical points is fitted best with a broken power-law and SMC or LMC curve. Fit with a MW curve, on the other hand, is successful with a single power-law spectrum. In the latter case the visual inspection reveals the model does not describe optical part of the SED very well. Best fit host extinction is lower than the one obtained from the X-shooter broadband fit.

The light curve of GRB\,120815A exhibits a smooth transition from shallow ($\alpha = 0.52 \pm 0.01$) to somewhat steeper ($\alpha = 0.86 \pm 0.03$) decay at $\approx 0.05$ days after the burst \citep{Kruhler2013}. Its behavior is achromatic in the NIR-to-X-rays, implying an absence of spectral break in this spectral regime.

\subsection{GRB\,130427A}

Photometric data used for normalization and photometric SED analysis were obtained from \citet{Perley2014}. The X-shooter SED alone can be fitted by a broken power-law and all three extinction curves. The broadband SED also requires a broken power-law shape. While the reduced $\chi^{2}$ implies the SMC or LMC curve provide a fit of a similar quality, a visual inspection shows that the model with the LMC-type dust fails to reproduce the data in the bluest X-shooter region. We thus prefer the SMC curve with low host extinction as the case best describing the real conditions. The break is required to occur within the observable X-shooter range. The difference in spectral slopes is $\Delta \beta = 0.31 \pm 0.05$. Fixing the spectral difference to $\Delta\beta = 0.5$ results in statistically worse, yet still acceptable fit. Visual inspection reveals that the X-shooter continuum in the latter case is not fitted that well. This fit also results in a shallower optical slope ($\beta_{\rm O} \sim 0.2$) and consequently higher extinction ($A_{\rm V} \sim 0.35$).

The broadband SED with photometric optical points is fitted best with a broken power-law---all three extinction curves can be used in the modeling of the optical SED. Results are in agreement with the X-shooter broadband SED fit and the SMC extinction curve.

The afterglow of GRB\,130427A comes with a rich multiwavelength data set and has been analysed in detail. It has been interpreted either as a combination of a forward and reverse shock afterglow contributions rising from a wind \citep{Laskar2013,Perley2014,Panaitescu2013} or ISM circumburst medium \citep{Maselli2014}. Regardless of the interpretation the cooling frequency lies at $\nu_{\rm C} > \nu_{\rm X}$ at the X-shooter epoch. Therefore the break we observe in the SED is not the cooling break, but may be due to the contribution of both forward and reverse shock to the emission \citep{Perley2014}.

\begin{table*}[t]
\small
\renewcommand{\arraystretch}{1.5}
\centering
\caption{Overview of the quality of fit results}
\begin{tabular}{lccccccccccr}
\hline\hline
       & \multicolumn{2}{c}{SMC} & & \multicolumn{2}{c}{LMC} & & \multicolumn{2}{c}{MW} & & \multicolumn{2}{c}{Best model}\\
\cline{2-3}
\cline{5-6}
\cline{8-9}
\cline{11-12}
GRB & $\left( \chi^{2}/{\rm dof}\right)_{\rm phot}$ & $\left( \chi^{2}/{\rm dof}\right)_{\rm spec}$ & & $\left( \chi^{2}/{\rm dof}\right)_{\rm phot}$ & $\left( \chi^{2}/{\rm dof}\right)_{\rm spec}$ & & $\left( \chi^{2}/{\rm dof}\right)_{\rm phot}$ & $\left( \chi^{2}/{\rm dof}\right)_{\rm spec}$ & & Phot & Spec \\
\hline
100219A & 3.8/10   & 41.7/30    & & 4.3/10 & 34.8/30       & & 6.9/10   & 44.2/30    & & MW & LMC\\
100418A & 11.2/12 &20.8/23     & & 10.7/12 & 20.2/23     & & 11.1/12 & 20.0/23    & & Any & SMC$^{\dagger}$\\
100814A & 48.1/33 &70.8/66     & & 47.8/33 & 71.0/66     & & 47.9/33 & 257/67     & & Any & SMC$^{\dagger}$\\
100901A & 15.4/26   &44.2/41     & & 14.8/26   & 160/47      & & 14.6/26   & 355/47      & & Any & SMC\\
120119A & 59.7/47 &194.1/81     & & 57.5/47 & 106.0/81   & & 79.5/47 &  1023/81   & & SMC/LMC & LMC\\
120815A & 21.3/21 & 26.0/47    & & 22.2/21 & 122.9/47    & & 20.0/22 & 353.1/27  & & Any & SMC\\
130427A & 62.6/64 &129.3/147 & & 62.5/64 & 130.0/147  & & 62.5/64 & 123.9/147 & & Any & SMC$^{\dagger}$\\
130603B & 11.5/9   &21.3/23     & & 11.1/9   & 20.9/23       & & 10.6/9   & 21.2/23    & & Any & SMC$^{\dagger}$\\
130606A$^{(a)}$ & 15.9/18 & 48.8/31    & &  16.0/18& 48.8/31       & & 16.0/18 &48.8/31    & & Any & Any\\
\hline
\end{tabular}
\tablefoot{Results of the fits to the broadband SEDs where photometry (phot) and X-shooter spectrum (spec) is used to build the NIR-to-UV part. Here we focus on the extinction curve component of the models. For each model we report the $\chi^{2}$/dof; here we consider only models where $\Delta \beta$ and $N_{\rm HI}$ are left as free parameters (see Table \ref{sedresults} for details and other models). The last two columns show whether a certain extinction curve is found to be preferential over the other two. In the X-shooter case, the better coverage of the blue part of the optical SED sometimes allows to distinguish between models even if their fits are of a similar quality according to the $\chi^2$ (see Section \ref{specvsphot}): such models are marked with $\dagger$.  \newline (a) The case of GRB\,130606A is found to have $A_{\rm V} \sim 0$.}
\label{tabchi}
\end{table*}

\subsection{GRB\,130603B}

This is the only short GRB in our sample. Photometric data used for normalization and photometric SED analysis were obtained from \citet{Postigo2014}. The X-shooter SED alone can be fitted by a broken power-law model and all three extinction curves. Broadband SED is modeled well with a broken power-law and all three extinction curves. Visual inspection shows the LMC and MW-type curves do not reproduce the blue part of the SED that well, therefore we prefer the SMC-type curve. A single power-law fails in reproducing the SED. In this case we included the $K$-band photometric observation in the fitting procedure: without this point the fitted pre-break slope would be too steep. However, we note that the NIR SED part cannot be modeled very well (see the residual plot in Figure \ref{figsed2}). The difference in spectral slopes is $\Delta \beta \sim 0.5$; fixing the spectral difference to $\Delta\beta = 0.5$ therefore does not change the results. A similar analysis has been done by \citet{Postigo2014}. They also find a high extinction of $A_{\rm V} \sim 0.9$, albeit a bit lower than we do ($A_{\rm V} \sim 1.2$). They find a position of the spectral break to be near $10^{16}$ Hz, while our best fit prefers a value of $\approx 0.8\times 10^{15}$ Hz.

The broadband SED with photometric optical points is fitted best with a broken power-law---all three extinction curves can be used in the modeling of the optical SED, though the MW curve seems to provide a slightly better fit than the other two curves. As in the X-shooter broadband fit, the spectral break is found in the optical region. However, best-modeled extinction $A_{\rm V}$ is found to be significantly lower. 

Optical afterglow observations of the GRB\,130603B are sparse. Optical and X-ray light curves appear to evolve achromatically after $\sim 0.25$ days, but are very different prior to that time \citep{Postigo2014}. During this time a gradual steepening from $\alpha \approx 1.4$ to $\alpha \approx 2.4$ in X-rays implies an occurrence of a jet break \citep{Fong2014}. The complicated light curve behavior prevents us to use the closure relations. We note that both \citet{Fong2014} and \citet{Postigo2014} find a spectral break between the optical and X-ray region, but at higher frequencies than we do.

\subsection{GRB\,130606A}

Photometric data used for normalization and photometric SED analysis were obtained from \citet{Tirado2013} and \citet{Afonso2013}. The fit to the X-shooter part of the SED alone is not very well constrained. Broadband SED is well represented by a power-law. No host extinction is necessary $(A_{\rm v} < 0.01)$. This is a high-redshift GRB ($z = 5.913$), therefore we had to include the Ly$\alpha$ absorption to constrain the UV slope. The value of the $\log N_{\rm HI} = 19.9 \pm 0.3$ that we obtain in the fit agrees with the one derived by \citet{Hartoog2014} (i.e., $\log N_{\rm HI} = 19.94 \pm 0.01$). Fixing the value to $\log N_{\rm HI} = 19.94$ therefore does not change the results.

For the photometric SED we use only filters not affected by host's Ly$\alpha$ and intergalactic medium absorption. Broadband SED with photometric optical points is fitted best with a single power-law and negligible extinction (while upper limits are quite high, clearly no extinction is necessary to model the SED). The results are in agreement with those obtained from the X-shooter broadband SED.

At the X-shooter epoch the light curve of GRB\,130606A decays achromatically with $\alpha \sim 1.9$ both at optical and X-ray wavelengths \citep{Tirado2013}. Our spectral slope and temporal decay index are consistent within the case of $\nu_{\rm C} > \nu_{\rm X}$ and wind circumburst environment.

\section{Discussion}
\label{discuss}

\subsection{Spectroscopic vs photometric SEDs}
\label{specvsphot}

To show the advantage of using spectroscopic SEDs in dust analysis we also modeled photometric SEDs\footnote{If not stated otherwise, we use the terms ``spectroscopic" and ``photometric" to refer to the broadband SEDs which include X-ray data.}, using the photometric data shown in Figures \ref{figsed1} - \ref{figsed2} (but adding $K$-band magnitudes, where available). Due to a poor sampling in blue SED parts, we did not use photometric points contaminated by Ly$\alpha$ absorption. While the difference in the spectral slopes $\Delta \beta$ in the case of photometric SED studies is usually fixed to $\Delta \beta = 0.5$, we leave it as a free parameter, because the same prescription was used in the spectroscopic SED study and therefore the comparison between the two analyses is more genuine. The results are reported in Table \ref{sedresults}.  

\begin{figure*}[!t]
\centering
\begin{tabular}{cc}
\includegraphics[scale=0.45]{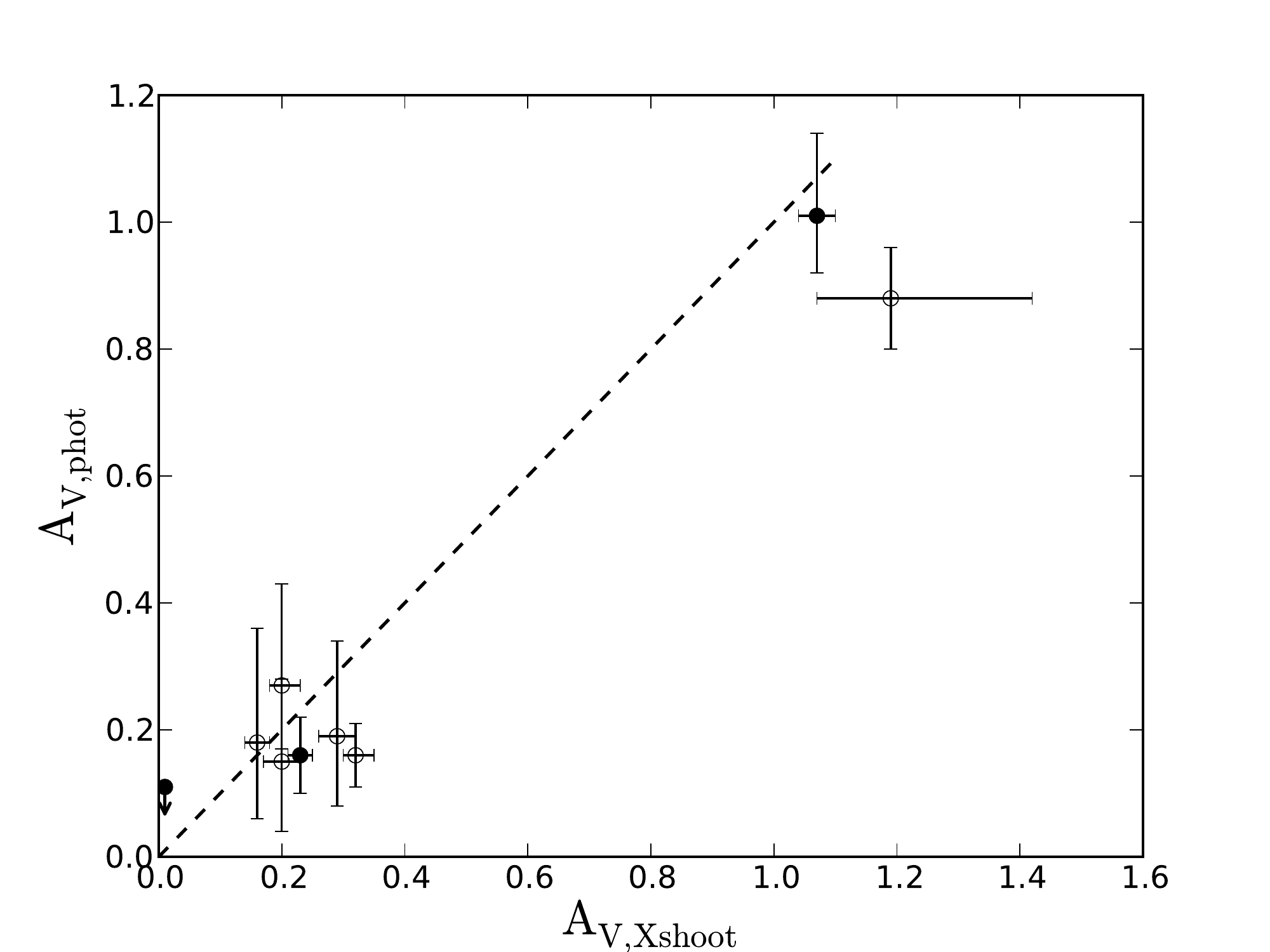}&
\includegraphics[scale=0.45]{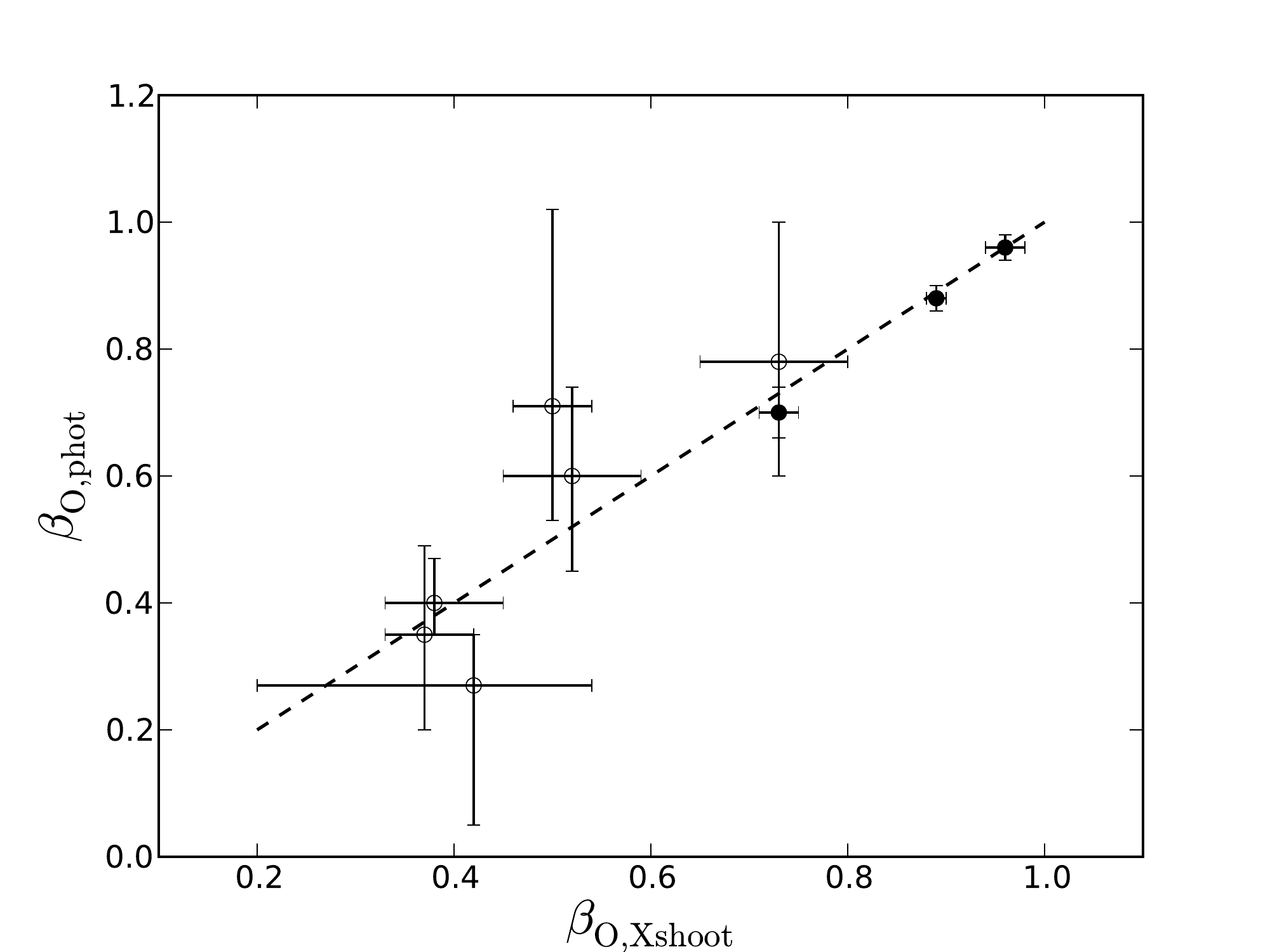}\\
 \end{tabular}
\caption{Comparison of best-fit parameters of extinction $A_{\rm V}$ (left) and optical spectral index $\beta_{\rm O}$ (right) obtained by fitting the broadband SED with photometric measurements and X-shooter spectra. Dashed lines represent relations $A_{\rm V,Xshoot} = A_{\rm V,phot}$ and $\beta_{\rm O,Xshoot} = \beta_{\rm O,phot}$. Filled and empty symbols represent the cases where the broadband SED is best described by a single or broken power-law, respectively.}
 \label{compare_av_beta}
\end{figure*} 

We find that the results of the modeling of photometric SEDs are similar in spectral shape to the results of modeling the spectroscopic SEDs. However,  photometric SEDs can usually be modeled by more than one extinction curve: in most cases we cannot strongly favor one of the models over the other two. Similar conclusions have been reached in other sample studies \citep[e.g.,][]{Covino2013}. The poor resolution of multicolor photometry is not sufficient to detect the smooth features in the SED which separate one extinction curve from another. This is in contrast to what is found in X-shooter SED analysis. In four cases of our sample one extinction curve is strongly preferred according to the $\chi^2$ statistics. In four additional cases, for which the best fit models with different extinction curves are of a similar quality, the better coverage of the blue part of the X-shooter SED allows us to distinguish the best one by visually inspecting the blue part of the SED, which is the most sensitive to the modeling. In these four cases the SMC-type extinction curve matched the blue data well, while for the other two curves either the 2175 \AA\, absorption was overestimated or the bluest SED part was overpredicted by the modeling (see Section \ref{results} for details). The ability to distinguish between the different extinction curves is summarized in Table \ref{tabchi}. In the spectroscopic analysis we were thus able to single out the best model for all cases but GRB\,130606A, whose line-of-sight lacks a notable extinction in the first place. 

Differences are also observed in the values of the best-fit parameters. Figure \ref{compare_av_beta} compares the values of extinction $A_{\rm V}$ and optical spectral index $\beta_{\rm O}$ obtained from both types of analysis. The extinction values seem to be systematically larger in the spectroscopic analysis, while optical spectral indices do not show any preferential deviation. Fixing the difference in spectral slope to $\Delta \beta = 0.5$, as it is usually done in these types of analyses \citep[e.g.,][]{Greiner2011,Covino2013}, does not reduce the differences significantly. Due to rather large errors in the values derived from the photometric analysis and small sample we cannot draw strong conclusions about possible trends in deviation. However, we emphasize that the photometric SEDs used in the analysis cover a very broad wavelength region. The analysis of SEDs with less covered spectral range, which are still often used in SED modeling, would result in even greater differences from the spectroscopic analysis.

The equivalent hydrogen column densities  were not constrained  very well by fitting only the X-ray part of the SED (see Table \ref{sedresults}). The reason for this is the low signal-to-noise of the X-ray spectrum: to minimize the error due to possible spectral evolution and uncertainties in temporal extrapolation, the X-ray SEDs were built from rather narrow time intervals. The broadband fit resulted in more constrained values of $N_{\rm H,X}$. We note that the spectral slopes in the X-ray are systematically steeper (although still within the error) in the broadband with respect to the X-ray-only fits. On the other hand, the fit of the X-shooter SED alone usually turned out very bad---only in three cases was such a fit both statistically acceptable (i.e., see Section \ref{modeling}) and resulted in similar parameter values as in the case of the broadband fit. Thus, independent of the quality (resolution) of the optical-to-NIR data, the SED modeling is secure only when the fit is performed on a broader band SED.

\begin{figure*}[t]
\centering
\begin{tabular}{cc}
\includegraphics[scale=0.45]{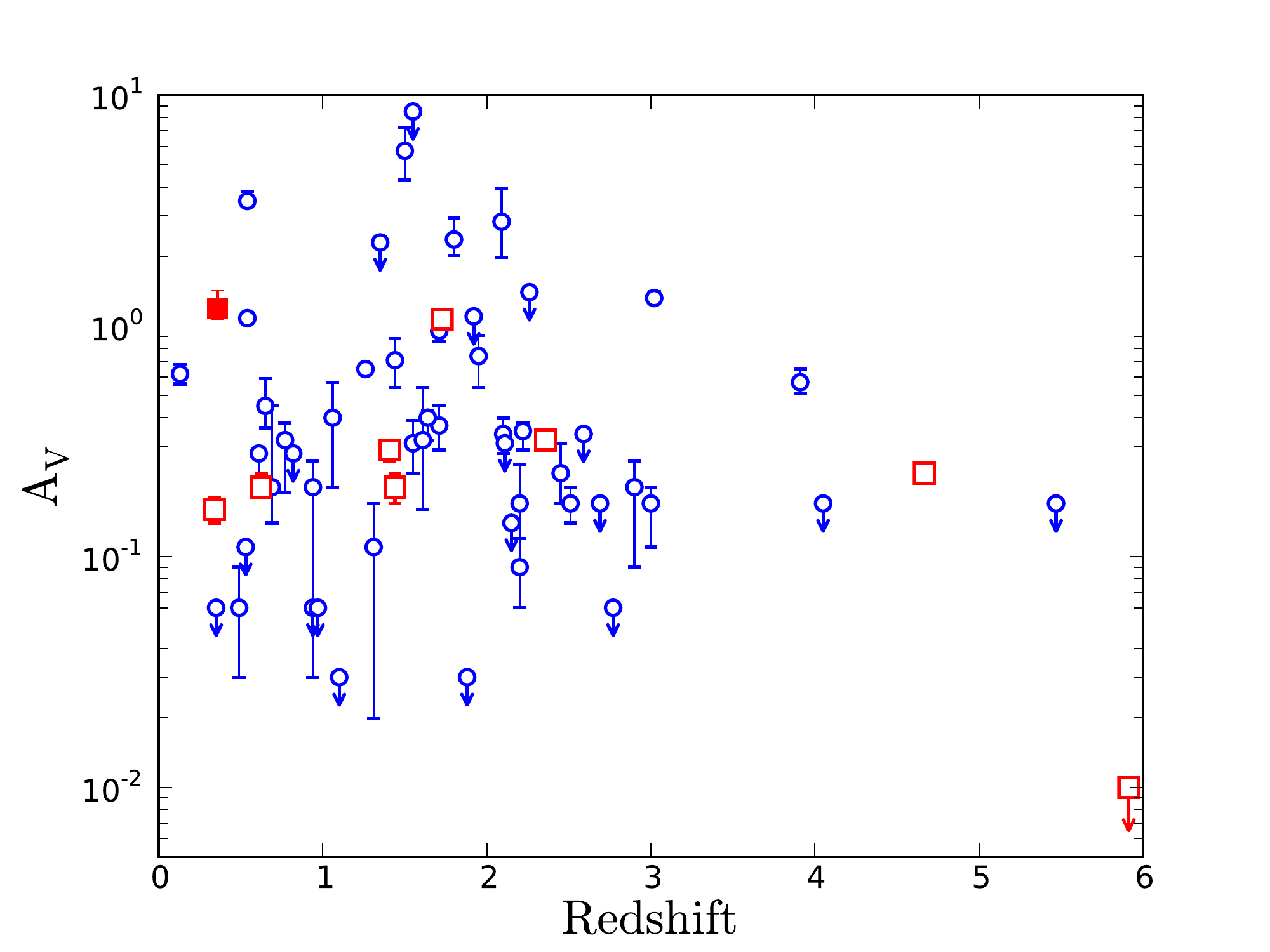}&
\includegraphics[scale=0.45]{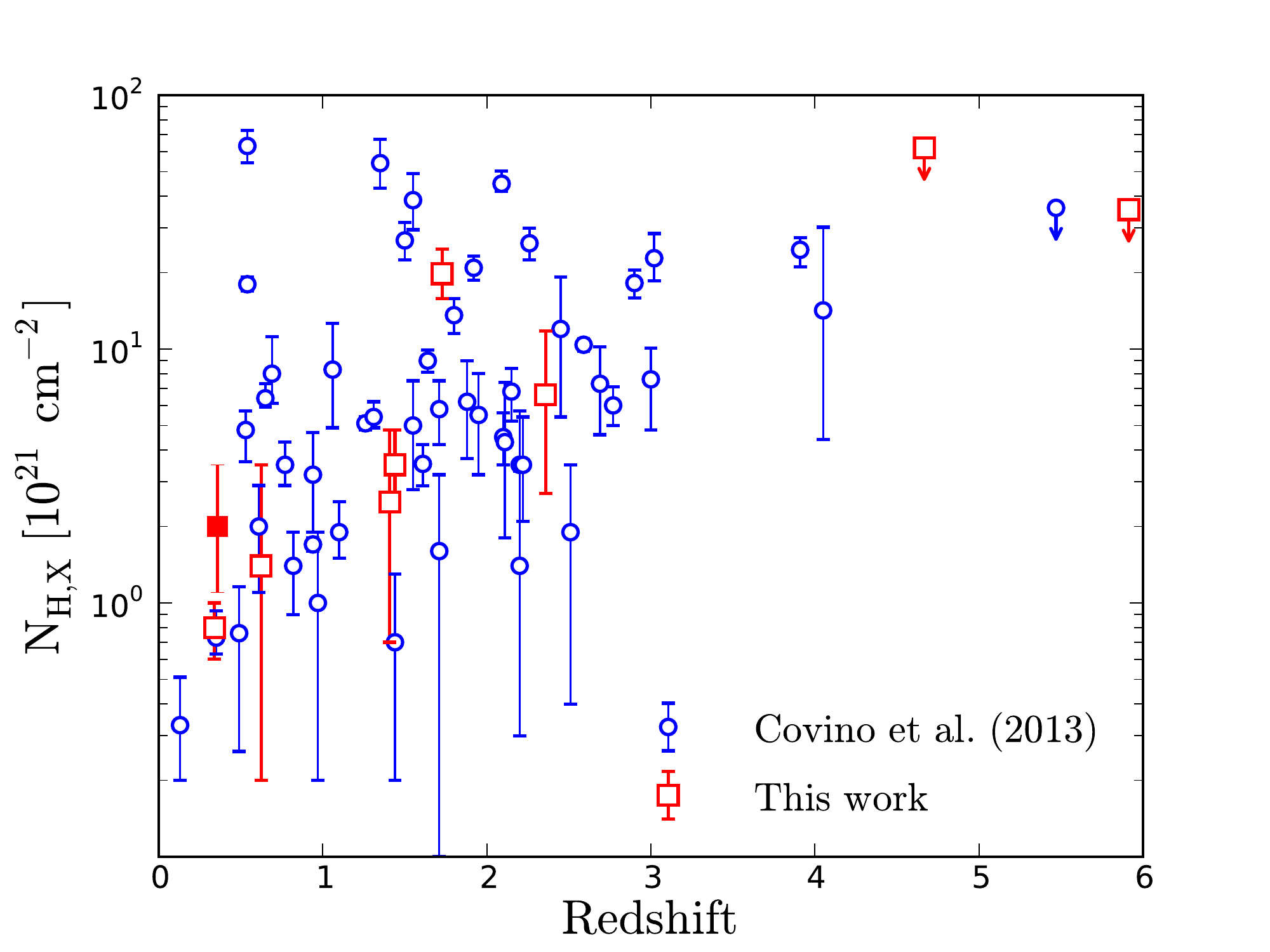}\\
 \end{tabular}
\caption{Extinction and hydrogen equivalent column density values as a function of redshift for our sample. Long GRBs are plotted with empty red squares while the short GRB\,130603B is given with a filled red square. Empty blue circles represent values obtained in a study of a complete sample of long GRBs \citep{Covino2013}.}
 \label{values}
\end{figure*} 

\subsection{Extinction and equivalent hydrogen column densities}
\label{ext_and_hyd}

The values of extinction ($0 \lesssim A_{\rm V} \lesssim 1.2$) and equivalent hydrogen column density ($0.1 \lesssim N_{\rm H,X} \left[ 10^{22} \mathrm{cm}^{-2}\right] \lesssim 2$; not including upper limits) that we find for our sample are similar to those found in GRB lines-of-sight \citep[e.g.,][]{Zafar2011,Greiner2011,Watson2012,Campana2012,Covino2013}. This is illustrated in Figure \ref{values} where we plot both quantities as a function of redshift and compare them to the {\it Swift} BAT6 complete GRB sample \citep{Salvaterra2012,Covino2013}. Most of the events are found to have low extinction ($A_{\rm V} \lesssim 0.3$). GRBs\,120119A and 130603B are moderately extinguished with $A_{\rm V} \sim 1.1$, which according to the complete {\it Swift} BAT6 sample places them into the top 20-25$\%$. The short GRB\,130603B has the highest extinction in the sample. Dust-to-gas ratios of our (long GRB) sample are low with respect to the ones of the Local Group, which is a well known result \citep{Stratta2004,Schady2010,Zafar2011}. The ratio also appears to increase with redshift--since the extinction does not evolve much with redshift (see Figure \ref{values}), the evolution of the ratio is a consequence of the lack of low $N_{\rm H,X}$ values at high redshifts \citep{Watson2013,Covino2013}. The reason for the preference of high $N_{\rm H,X}$ values at high redshifts is not yet clear. \citet{Campana2012} claims that for high-redshift GRBs the absorption by intervening systems in the GRB line-of-sight contributes much to the measured $N_{\rm H,X}$. Alternatively, \citet{Watson2013} claims that the absorption is intrinsic to the GRB environment and that the evolution of the (metal) gas column density reflects the evolution of cosmic metallicity. As already noted by \citet{Postigo2014}, the dust-to-gas ratio of the short GRB\,130603B is consistent with the Galactic, indicating that the explosion site for this event differs from a typical long GRB site.

\subsection{Extinction curves}

We find that the dust properties preferred for the GRB sight-lines are those of the SMC type, with six afterglow SEDs best fitted by the corresponding averaged extinction curve. Two are best described by the LMC curve and one (GRB\,130606A) is found to have $A_{\rm V} \sim 0$. While the preference for the SMC-type of dust has been already observed in the sample studies using photometric SEDs \citep[e.g.][]{Kann2010,Greiner2011,Covino2013} and the spectroscopic study of Z11, it is still interesting that the well sampled and broadband X-shooter SEDs can be modeled with this average extinction curve that well. The lack of the 2175 $\mathrm{\AA}$ bump in all but one event (see below) is not surprising: Z11 found that the events with a notable 2175 $\mathrm{\AA}$ feature all have rather high extinction values and that the preference for the SMC-type dust can be attributed to the observed GRBs being biased towards low extinction lines-of-sight. Indeed, most of our events have low measured extinction. If the total-to-selective extinction $R_{\rm V}$ in the lines-of-sight in GRB hosts were larger than in the three assumed extinction curves, the curves would be flatter and the derived extinction higher. Such extinction could occur if dust were being destroyed by a GRB and grains of smaller size are preferentially destroyed \citep{Waxman2000}. However, while such dust may have been found in some events \citep{Perley2008,Liang2010}, as already emphasized in Section \ref{modeling}, we find flat extinction curve to be inadequate to describe the dust in all lines-of-sight of our GRB sample.

GRB\,120119A is the only event in our sample with a clear 2175 $\mathrm{\AA}$ absorption bump (see Figure \ref{sed120119A}), the feature observed also in a few other afterglow spectra \citep{Eliasdottir2009,Prochaska2009,Zafar2012,Fynbo2014}. In our analysis the LMC template provided the best fit to the data for this GRB. Still, it is clear from Figure \ref{sed120119A} that the LMC overpredicts the strength of the 2175 $\mathrm{\AA}$ bump. The failure to find a good model is not completely unexpected, since we are using merely average extinction curves. Different lines-of-sight in our Galaxy or in the two Magellanic Clouds have different extinction properties - there are known lines-of-sight in the Galaxy having SMC-type dust and vice versa (e.g., see \citealt{Eliasdottir2009} for review). The analysis thus calls for a more detailed extinction model, like the one introduced by \citet{Fitzpatrick2007}, where the strength of the bump as well as the UV extinction slope are free parameters of the model. The use of more general extinction curves in the X-shooter SED analysis will be presented in a separate work (Zafar et al. in prep.). GRB\,120119A is a perfect example with excellent data both in optical and X-ray frequency range that clearly shows the power of the X-shooter data---a photometric SED can hardly differentiate between the three models, while the X-shooter data allow us to extract much more detailed information about the dust.

\subsection{Neutral hydrogen column densities}

Four GRBs in the sample are at a redshift $z \gtrsim$ 1.7 for which the red wings of the Ly$\alpha$ absorption line enter the X-shooter observational window and influence the SED shape. The modeled hydrogen column density values for these bursts are given in Table \ref{bestfit}. In two cases our values agree with the ones derived from normalized spectra (i.e., GRB\,100219A - \citealt{Thoene2013}; GRB\,130606A - \citealt{Hartoog2014}). For the other two we derive values that are slightly (GRB\,120815A - \citealt{Kruhler2013}) or significantly (GRB\,120119A - Vreeswijk et al. 2014 in prep) higher from those in the literature. In the case of the GRB\,120815A, the Ly$\alpha$ line is contaminated by vibrationally excited H$_{2}^{*}$ lines that form a continuum around this spectral region and cause the line to appear stronger \citep{Kruhler2013}. The contribution of the H$_{2}^{*}$ absorption has already been subtracted from the value of $\log N_{\rm HI} = 21.95 \pm 0.15$, reported by \citet{Kruhler2013}. Not taking the molecular hydrogen into account, they measure $\log N_{\rm HI} \sim 22.1$, closer to our value. 

\begin{figure}[t]
\centering
\includegraphics[scale=0.47]{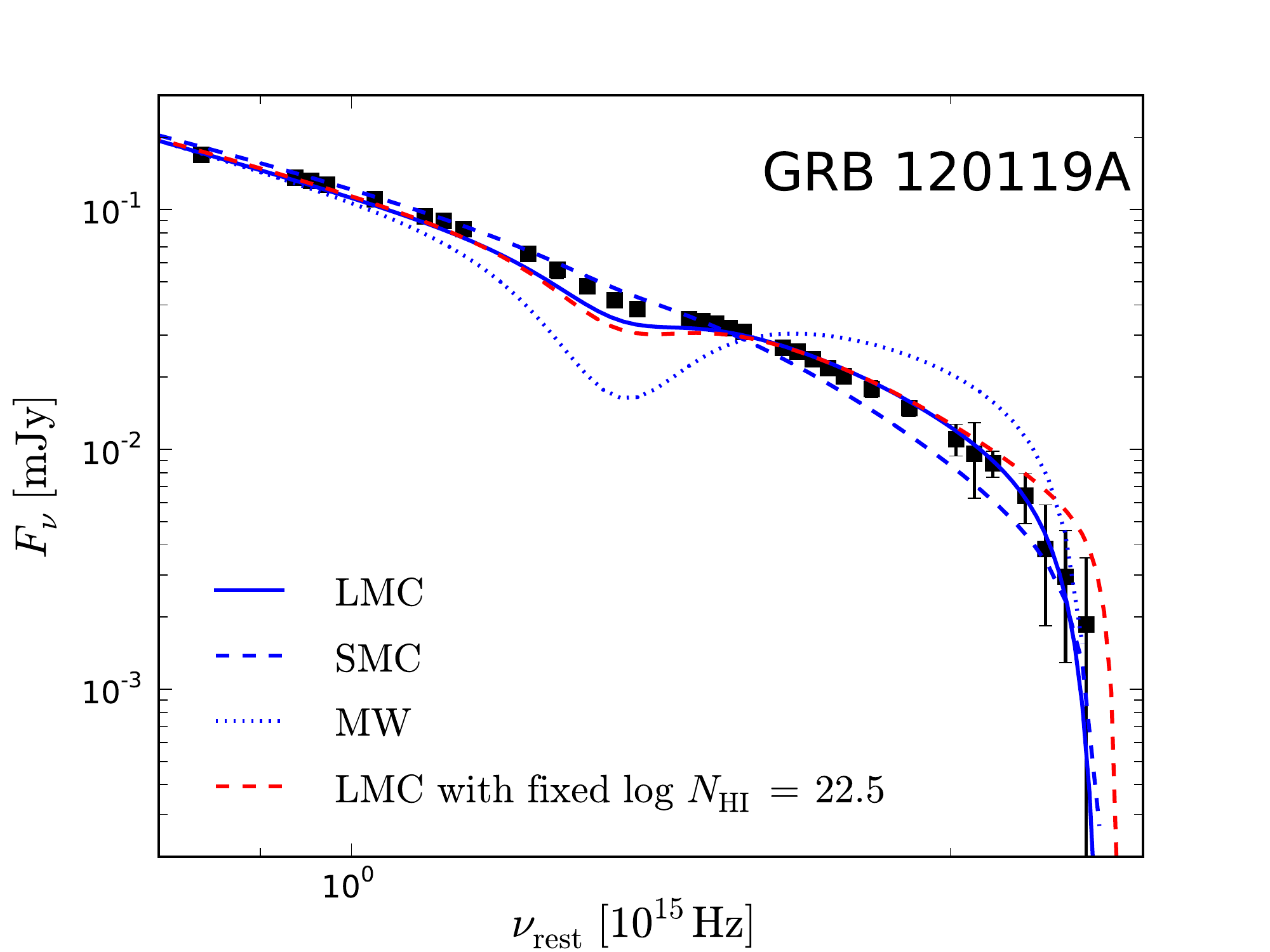}
\caption{Zoomed blue X-shooter part of the broadband SED fit of GRB\,120119A. Several models are shown, indicating their failure to fully reproduce the data. }
 \label{sed120119A}
\end{figure}

There are several possible reasons for the further discrepancy. Firstly, we normalized the spectra of the four GRB afterglows and fitted the Ly$\alpha$ lines: our best-fit $N_{\rm HI}$ match very well with the values from the literature. Secondly, our spectra are heavily binned. To check the dependency of the results on the bin size, we redid the modeling of the X-shooter SED part but with smaller bin widths in the region around the line: the obtained value of the $\log N_{\rm HI}$ does not change significantly. We also fixed the $\log N_{\rm HI}$ to the values measured from the normalized spectra (see Table \ref{sedresults} and Section \ref{results}). The results did not change significantly for GRBs 100219A and 130606A. In the case of GRB\,120815A the fit statistics was worse, but otherwise the parameter values did not change significantly. The case of GRB\,120119A remains puzzling. The value of $\log N_{\rm HI} = 23.4 \pm 0.2$ found for GRB\,120119A seems to be unrealistically high. While GRB lines-of-sight are characterized by generally high hydrogen column densities \citep{Fynbo2009, Thoene2013}, the measured values have never surpassed $\log N_{\rm HI} = 23.0$. Indeed, the column density measured from the normalized spectrum is found to be $22.5 \pm 0.3$ (Vreeswijk et al. in prep), but fixing this value in the fitting procedure cannot reproduce the blue part of the SED (see Figure \ref{sed120119A}). The main reason for the discrepancy is probably due to a low signal-to-noise of the spectral region in which the line lies (i.e., the bluest part of the spectrum). The uncertainty of our result is further enhanced by the shape of the extinction curves that in this case have proved to be inadequate to model the extinction.

\begin{figure}[t]
\centering
\includegraphics[scale=0.47]{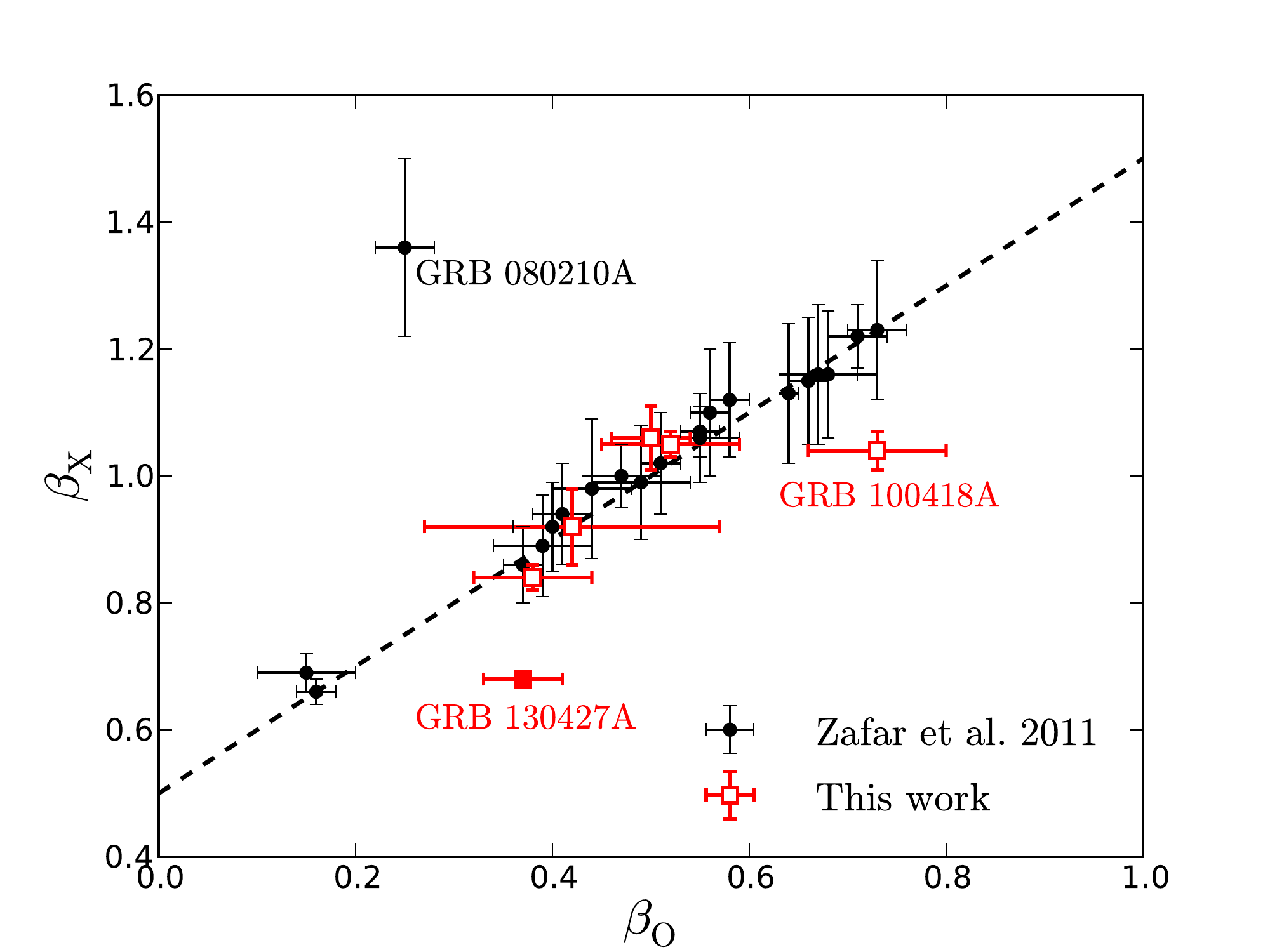}
\caption{Comparison of optical and X-ray spectral slopes for the cases where spectral break is found to lie in the analysed spectral range. Our values (empty red squares) are compared to the sample of \citet{Zafar2011} (filled black circles). Dashed line represents the relation $\beta_{\rm X} = \beta_{\rm O} + 0.5$, corresponding to a theoretical prediction in the case the observed break is the cooling break. The data point corresponding to GRB\,130427A, for which the break probably does not correspond to a cooling break, is plotted with filled square.}
 \label{deltabreak}
\end{figure}

\subsection{Standard afterglow model}

While this is not the main goal of the paper, here we present a brief discussion of our results as analysed with respect to the standard afterglow theory. 

A spectral break is necessary to model the SEDs of six GRBs in our sample: GRB\,100418A, 100814A, 100901A, 100815A, 130427A and 130603B. An occurrence of a spectral break in the SED is not surprising: analysing a sample of $\sim 40$ afterglows, Z11 find that $\sim 60\%$ of SEDs have a spectral break lying somewhere between the optical and X-ray frequencies. Spectral slopes for the six GRBs, together with data of the sample presented by Z11, are plotted in Figure \ref{deltabreak}. In four of our cases, the difference between pre- and post-break slopes of our GRBs is in agreement within errors with theoretically predicted value of  $\Delta \beta = 0.5$, confirming that both optical and X-ray emission have a synchrotron origin and are produced by the same mechanism - similar conclusions were reached by Z11. Nevertheless, we find two exceptions to the expected behavior.  The SEDs of GRBs 100418A and 130427A are found to have a break with much shallower $\Delta \beta \sim 0.3$. The afterglow of the GRB\,130427A has been interpreted either as a combination of a forward and reverse afterglow contribution rising from a wind \citep{Laskar2013,Perley2014,Panaitescu2013} or ISM circumburst medium \citep{Maselli2014}. Regardless of the interpretation, the cooling frequency is found to lie above the X-ray frequency at the X-shooter epoch, suggesting that the break in the SED is not due to the passing of the cooling frequency through the optical SED. This is supported by the fact that the measured change in spectral slopes differs from the one predicted by the standard theory for the cooling break. The break we find may thus be due to the contribution of both forward and reverse shock to the emission \citep{Perley2014}. The case of GRB\,100418A is harder to explain. Its optical afterglow is characterized by a strong bump which is not seen in the X-rays - our spectrum is taken when the bump is still strong. A contribution of multiple components in the optical may be the cause of the shallow $\Delta \beta =  0.31 \pm 0.09$. Even greater discrepancy was found for GRB\,080210A by Z11 and \citet{DeCia2011}, for which the spectral difference is $\Delta \beta \sim 1$. However we recall that we used a sharp spectral break in the fitting procedure (see Section \ref{modeling}). If the actual spectral transition is somewhat milder, occurring over several orders of magnitudes, the modeled spectral difference would be underestimated by using a sharp break. We also point to rather large uncertainties of optical spectral indices. This is because spectral breaks lie within or near the optical spectral region and consequently the spectral range needed to constrain the optical slopes is quite narrow.

We briefly checked the available afterglow light curves of the GRBs in our sample to see whether the relations between spectral and temporal indices are in agreement with theoretical expectations in the case of different physical models \citep[e.g., Table 1 in ][]{Racusin2009}. A detailed account on the consistency with closure relations for each burst is given in Section \ref{results}. While some cases can indeed be explained within one of the models' predictions (e.g., 100814A and 130606A), most of the events are not consistent with closure relations. This is a known problem observed in other sample studies \citep[e.g., Z11, ][]{Zaninoni2013} and probably reflects our ignorance of the hydrodynamic evolution of the outflow and the behavior of the central engine. 

\section{Conclusions}
\label{conclude}

We have presented a detailed SED analysis of a sample of GRB afterglows, observed with the VLT/X-shooter spectrograph. Our aim was to model broadband afterglow SEDs, composed of the NIR-to-ultraviolet X-shooter spectra and {\it Swift}/XRT X-ray observations, to infer dust properties along the lines-of-sight in GRB host galaxies. Only spectra with trustworthy flux calibration (i.e., for which the spectral slope matches with the one built from photometric measurements) were used in the analysis. The sample consists of nine GRBs: eight GRBs belong to the long and one to the short class. 

The values of extinction ($0 \lesssim A_{\rm V} \lesssim 1.2$) and equivalent hydrogen column density ($0.1 \lesssim N_{\rm H,X} \left[ 10^{22} \mathrm{cm}^{-2}\right] \lesssim 2$) which we measure are commonly found in GRB sightlines \citep[e.g.,][]{Covino2013}. Six GRBs in the sample require a spectral break in the modeled spectral region. Interestingly, four breaks occur in the region which is covered by X-shooter. Most of the events are best described by the SMC-type of dust.

We redid the modeling using the broadband SEDs with the NIR/UV photometric measurements in order to see whether there are any differences in using the two data sets in the analysis. The derived values of the extinction $A_{\rm V}$ and spectral slope, obtained through spectroscopic and photometric SED analysis, for individual events can differ significantly, though no apparent trend in the differences is observed. More importantly, the spectroscopic data, especially with their excellent coverage of the blue SED part, can help us to constrain the dust properties (e.g., the extinction curve) much better than the photometric measurements. In addition, we show that independently of the quality (resolution) of the optical-to-NIR data, the SED modeling is secure only when the fit is performed on a SED covering broader spectral region (in our case, including the X-rays).

We have demonstrated that spectra, obtained with X-shooter, can be successfully used for a detailed SED analysis. X-shooter spectra with good flux calibration can be used to constrain the extinction curves and can therefore improve our understanding of dust in the high redshift Universe. The study will be expanded, when more photometric observations of afterglows become available in the future.

\begin{acknowledgements}
AdUP acknowledges support from a Ram\'on y Cajal fellowship. The Dark Cosmology Centre is funded by the Danish National Research Foundation. This work made use of data supplied by the UK Swift Science Data Centre at the University of Leicester.
\end{acknowledgements}


\bibliographystyle{aa}
\bibliography{xshoot_ext_bib}

\appendix

\section{Fitting details}
\label{sec:details}

\onecolumn

\begin{longtab}
\renewcommand{\arraystretch}{1.5}
\begin{landscape}
\tiny
\begin{longtable}{lcccccccccccccccr}
\caption{Fitting results. We do not report errors in cases with high values of reduced $\chi^{2}$, for which the procedure for measuring errors could not be safely applied. For each GRB we first fit only the optical (X-shooter) part using different extinction profiles. Then we repeat the fitting by including the X-ray data into the fit and compare the results. For each modelled extinction curve we report the best result, i.e., power-law or broken power-law model. Since including the neutral hydrogen column density as a free parameter in the fit may represent an uncertain element that could significantly affect the results, we also show the fits in which $N_{\rm HI}$ is taken from the literature - these are marked with $f$. In addition, we report the best fit to the XRT SED and to the broadband SED with photometric optical data points. The latter are marked with letter $p$. XRT-only SEDs are always fitted best with a single power-law model.}\label{sedresults}\\
\hline\hline
       &                 &\multicolumn{7}{c}{X-shooter only/XRT only/broadband photometric} & & \multicolumn{7}{c}{X-shooter $\&$ XRT}  \\
\cline{3-9}
\cline{11-17}
GRB & Extinction & $N_{\rm H,X}$ & $\beta_{1}$ & $\beta_{2}$ & $\nu_{\rm break}^{(a)}$ & $\log N_{\rm HI}$(Ly$\alpha$) & Av &  $\chi^{2}$/dof & & $N_{\rm H}$                       & $\beta_{1}$ & $\beta_{2}$ & $\nu_{\rm break}^{(a)}$ & $\log N_{\rm HI}$(Ly$\alpha$) & Av &  $\chi^{2}$/dof \\
       &                 &   $[10^{22} {\rm cm}^{-2}]$   &       &                    & $[10^{15}$ Hz]               &  $\log$ cm$^{-2}$                    &       &                          & & $[10^{22} {\rm cm}^{-2}]$  &                    &                    & $[10^{15}$ Hz]             & $\log$ cm$^{-2}$                        &     &\\
\hline
\endfirsthead
\caption{continued.}\\
\hline\hline
       &                 &\multicolumn{7}{c}{X-shooter only/XRT only/broadband photometric} & & \multicolumn{7}{c}{X-shooter $\&$ XRT}  \\
       \cline{3-9}
\cline{11-17}
GRB & Extinction & $N_{\rm H,X}$ & $\beta_{1}$ & $\beta_{2}$ & $\nu_{\rm break}^{(a)}$ & $\log N_{\rm HI}$(Ly$\alpha$) & Av &  $\chi^{2}$/dof & & $N_{\rm H}$                       & $\beta_{1}$ & $\beta_{2}$ & $\nu_{\rm break}^{(a)}$ & $\log N_{\rm HI}$(Ly$\alpha$) & Av &  $\chi^{2}$/dof \\
       &                 &   $[10^{22} {\rm cm}^{-2}]$   &       &                    & $[10^{15}$ Hz]               &  $\log$ cm$^{-2}$                    &       &                          & & $[10^{22} {\rm cm}^{-2}]$  &                    &                    & $[10^{15}$ Hz]             & $\log$ cm$^{-2}$                        &     &\\
\hline
\endhead
\hline
\endfoot
100219A & SMC & & 0.46$_{-0.37}^{+0.29}$  & & & 20.9$_{-0.1}^{+0.1}$ & 0.21$_{-0.07}^{+0.09}$ & 37.0/21 & & < 5.9 & 0.71$_{-0.02}^{+0.02}$ & & & 21.0$_{-0.1}^{+0.1}$ & 0.15$_{-0.02}^{+0.02}$ & 41.7/30 \\
               & LMC & & 0.49$_{-0.36}^{+0.29}$  & & & 20.9$_{-0.1}^{+0.1}$ & 0.32$_{-0.08}^{+0.09}$ & 30.1/21 & & < 6.2 & 0.73$_{-0.02}^{+0.02}$ & & & 21.0$_{-0.1}^{+0.1}$ & 0.23$_{-0.02}^{+0.02}$ & 34.8/30 \\ 
               & MW  & & 0.73                                & & & 20.9                           & 0.43                                & 44.0/21 & & < 4.3 & 0.77$_{-0.02}^{+0.02}$ & & & 20.9$_{-0.2}^{+0.1}$ & 0.40$_{-0.02}^{+0.03}$ & 44.2/30 \\ 
               & SMC$^{f}$ & & 0.53                       & & & 21.14                         & 0.18                                & 44.5/22 & & < 5.8 & 0.71$_{-0.02}^{+0.02}$ & & & 21.14 & 0.14$_{-0.02}^{+0.02}$ & 47.7/31 \\
               & LMC$^{f}$ & & 0.58$_{-0.20}^{+0.29}$  & & & 21.14                         & 0.27$_{-0.12}^{+0.07}$ & 39.4/22 & & < 6.1 & 0.72$_{-0.02}^{+0.02}$ & & & 21.14 & 0.22$_{-0.02}^{+0.02}$ & 42.4/31 \\ 
               & MW$^{f}$  & & 0.99$_{-0.11}^{+0.13}$  & & & 21.14                         & 0.23$_{-0.10}^{+0.09}$ & 42.3/22 & & < 5.7 & 0.76$_{-0.02}^{+0.02}$ & & & 21.14 & 0.39$_{-0.02}^{+0.03}$ & 55.5/31 \\ 
               &  XRT &$<$ 5.9 & 0.57$_{-0.27}^{+0.39}$ & & & & & 2.2/7 & & & & & & & & \\
               & SMC$^{p}$ &  < 5.5 & 0.69$_{-0.03}^{+0.03}$ & & &                                   & 0.11$_{-0.04}^{+0.04}$ & 3.8/10 & & & & & & & &\\
               & LMC$^{p}$ &  < 5.6 & 0.70$_{-0.04}^{+0.04}$ & & &                                   & 0.16$_{-0.06}^{+0.06}$ & 4.3/10 & & & & & & & &\\ \vspace{0.25cm}
               & MW$^{p}$ &  < 5.8 & 0.72$_{-0.05}^{+0.05}$ & & &                                   & 0.25$_{-0.11}^{+0.11}$ & 6.9/10 & & & & & & & &\\

100418A 	& SMC & & 0.50$_{-0.17}^{+0.11}$  &  &  & & 0.52$_{-0.09}^{+0.11}$ & 24.0/16  & & 0.14$_{-0.12}^{+0.21}$ & 0.73$_{-0.08}^{+0.07}$ & 1.04$_{-0.03}^{+0.03}$ & 0.6$_{-0.1}^{+0.3}$ & & 0.20$_{-0.02}^{+0.03}$ & 20.8/23\\
				& LMC & & 0.50$_{-0.16}^{+0.12}$ &   &  & & 0.56$_{-0.09}^{+0.11}$ & 33.8/16  & & 0.14$_{-0.12}^{+0.21}$ & 0.72$_{-0.08}^{+0.07}$ & 1.05$_{-0.03}^{+0.03}$ & 0.6$_{-0.1}^{+0.3}$ & & 0.22$_{-0.02}^{+0.03}$ & 20.2/23\\
				& MW  & & 0.65$_{-0.10}^{+0.13}$ &   & & & 0.44$_{-0.10}^{+0.08}$ & 45.8/16  & & 0.13$_{-0.12}^{+0.20}$ &  0.73$_{-0.08}^{+0.07}$ & 1.05$_{-0.03}^{+0.03}$ & 0.6$_{-0.1}^{+0.3}$ & & 0.21$_{-0.02}^{+0.03}$ & 20.0/23\\
                & XRT  &$<$ 0.3 & 0.85$_{-0.31}^{+0.44}$ & & & & & 6.3/7 & & & & &  & & & \\
			    & SMC$^{p}$ & 0.13$_{-0.11}^{+0.20}$ & 0.78$_{-0.18}^{+0.22}$ & 1.01$_{-0.06}^{+0.11}$ & 1.0$_{-0.5}^{+15.1}$ &                                   & 0.27$_{-0.10}^{+0.16}$ & 11.2/12 & & & & & & & &\\
				& LMC$^{p}$ & 0.13$_{-0.11}^{+0.20}$ & 0.77$_{-0.20}^{+0.26}$ & 1.02$_{-0.06}^{+0.14}$ & 1.1$_{-0.4}^{+14.8}$ &                                   & 0.30$_{-0.11}^{+0.24}$ & 10.7/12 & & & & & & & &\\\vspace{0.25cm}	
				& MW$^{p}$  & 0.14$_{-0.11}^{+0.21}$ & 0.79$_{-0.20}^{+0.25}$ & 1.02$_{-0.06}^{+0.15}$ & 1.1$_{-0.6}^{+18.2}$ &                                   & 0.29$_{-0.11}^{+0.27}$ & 11.1/12 & & & & & & & &\\

100814A  & SMC & & 0.52$_{-0.07}^{+0.06}$ & & & & 0.20$_{-0.03}^{+0.02}$ & 25.5/38   & &  0.35$_{-0.11}^{+0.13}$ & 0.52$_{-0.07}^{+0.07}$ & 1.05$_{-0.02}^{+0.02}$ & 2.3$_{-0.1}^{+3.8}$ & & 0.20$_{-0.03}^{+0.03}$ & 70.8/66\\
               & LMC & & 0.49                               & &  & & 0.29                               & 111.7/38  & &  0.30$_{-0.10}^{+0.12}$  & 0.58$_{-0.05}^{+0.05}$ & 1.99$_{-0.01}^{+0.01}$ & 1.3$_{-0.1}^{+0.1}$ & & 0.16$_{-0.02}^{+0.02}$ & 71.0/66\\
               & MW  & & 0.99                               & &  &  & 0.0                                & 187.1/38  & & 0.19                                & 0.94                               &                                       &                                       &  & 0.02                              & 257/68\\
               & XRT &0.21$_{-0.16}^{+0.18}$ & 0.87$_{-0.16}^{+0.17}$ & & & & & 42.8/28 & & & & & & & & \\
               & SMC$^{p}$ & 0.37$_{-0.11}^{+0.13}$ & 0.65$_{-0.13}^{+0.11}$ & 0.86$_{-0.04}^{+0.06}$ & 1.7$_{-0.7}^{+9.6}$ &                                   & 0.10$_{-0.08}^{+0.10}$ & 48.1/33 & & & & & & & &\\
			    & LMC$^{p}$ & 0.37$_{-0.11}^{+0.14}$ & 0.60$_{-0.15}^{+0.14}$ & 0.88$_{-0.05}^{+0.06}$ & 1.7$_{-0.7}^{+7.4}$ &                                   & 0.15$_{-0.11}^{+0.13}$ & 47.8/33 & & & & & & & &\\\vspace{0.25cm}
			    & MW$^{p}$  & 0.37$_{-0.11}^{+0.13}$ & 0.60$_{-0.15}^{+0.15}$ & 0.88$_{-0.06}^{+0.05}$ & 1.7$_{-0.8}^{+7.8}$ &                                   & 0.16$_{-0.12}^{+0.13}$ & 47.9/33 & & & & & & & & \\

100901A & SMC  & & 0.26$_{-0.10}^{+0.17}$ &  &  & & 0.39$_{-0.07}^{+0.04}$ & 18.8/18  & & 0.25$_{-0.18}^{+0.23}$ & 0.50$_{-0.04}^{+0.04}$ & 1.06$_{-0.06}^{+0.05}$ & 5.8$_{-3.2}^{+8.8}$ & & 0.29$_{-0.03}^{+0.03}$ & 44.2/41\\
               & LMC  & & 0.81                               &  &  & & 0.18                               & 147.2/18 & & 0.26$_{-0.18}^{0.24}$ & 0.81                               & 1.10                               & 150.8                       & & 0.21  & 160.0/41\\
               & MW   & & 1.18                               &  &  & & 0.0                                 & 169.2/18 & & < 1.5                            & 0.93                               & 1.20                               & 15000                      & & 0.15  & 355.0/41\\
               & XRT &$<$ 0.6 & 1.10$_{-0.23}^{+0.26}$ & & & & & 18.9/23 & & & & & & & & \\
               & SMC$^{p}$ & 0.60$_{-0.44}^{+0.35}$ & 0.71$_{-0.18}^{+0.31}$ & 1.23$_{-0.06}^{+0.07}$ & 34.9$_{-29.1}^{+413}$ &                                   & 0.19$_{-0.11}^{+0.15}$ & 15.4/26 & & & & & & & &\\
				& LMC$^{p}$ & 0.61$_{-0.44}^{+0.35}$ & 0.73$_{-0.17}^{+0.28}$ & 1.23$_{-0.05}^{+0.07}$ & 29.0$_{-23.3}^{+623}$ &                                   & 0.28$_{-0.16}^{+0.22}$ & 14.8/26 & & & & & & & &\\\vspace{0.25cm}
				& MW$^{p}$  & 0.58$_{-0.38}^{+0.54}$ & 0.72$_{-0.14}^{+0.33}$ & 1.20$_{-0.05}^{+0.06}$ & 17.5$_{-17.0}^{+450}$ &                                   & 0.38$_{-0.23}^{+0.27}$ & 14.6/26 & & & & & & & &\\

120119A & SMC & & 1.61$_{-0.11}^{+0.15}$ &  & & 23.1$_{-0.4}^{+0.3}$ & 0.50$_{-0.03}^{+0.03}$ & 58.1/34   &  &  1.78                                & 0.86                               & & & 23.7                           & 0.88                               & 194.1/81\\      
               & LMC & & 1.50$_{-0.22}^{+0.14}$ &  & & 23.3$_{-0.2}^{+0.2}$  & 0.67$_{-0.08}^{+0.09}$ & 62.9/34  &  &   1.98$_{-0.40}^{+0.50}$ & 0.89$_{-0.01}^{+0.01}$ & & & 23.4$_{-0.2}^{+0.2}$ & 1.07$_{-0.03}^{+0.03}$ & 106.0/81\\
               & MW  & & 2.57                               &  & & 23.8                            & 0.01                               & 311.0/34 & &  1.75									& 0.92                               & & & 23.8                           & 1.26                              & 1023/81\\
				& SMC$^{f}$ & & 1.57 &  & & 22.5 & 0.53 & 87.1/35 &  &  1.79                                & 0.86                               & & & 22.5                           & 0.88                               & 182.5/82\\      
               & LMC$^{f}$ & & 1.18 &  & & 22.5  & 0.92 & 96.9/35&  &   2.01$_{-0.41}^{+0.50}$ & 0.89$_{-0.01}^{+0.01}$ & & & 22.5                           & 1.10$_{-0.03}^{+0.03}$ & 126.3/82\\
               & MW$^{f}$  & & 2.60 &  & & 22.5  & 0.01 & 331.0/35& &  1.75									& 0.93                               & & & 22.5                           & 1.31                              & 1268/82\\             
               & XRT &1.15$_{-0.40}^{+0.51}$ & 0.60$_{-0.12}^{+0.13}$ & & & & & 31.6/45 & & & & & & & & \\
               & SMC$^{p}$ & 2.07$_{-0.42}^{+0.52}$ & 0.86$_{-0.02}^{+0.02}$ & & &                                   & 0.93$_{-0.07}^{+0.10}$ & 59.7/47 & & & & & & & &\\
               & LMC$^{p}$ & 2.11$_{-0.43}^{+0.53}$ & 0.88$_{-0.02}^{+0.02}$ & &  &                                    & 1.01$_{-0.09}^{+0.13}$ & 57.5/47 & & & & & & & &\\\vspace{0.25cm}
               & MW$^{p}$  & 2.11$_{-0.43}^{+0.53}$ & 0.88$_{-0.02}^{+0.02}$ & &  &                                   & 1.03$_{-0.08}^{+0.12}$ & 79.5/47 & & & & & & & &\\

120815A & SMC	& & 0.49$_{-0.29}^{+0.23}$ &1.11$_{-0.40}^{+0.60}$   & 1.4$_{-0.1}^{+0.1}$  & 22.3$_{-0.3}^{+0.2}$ & 0.26$_{-0.17}^{+0.15}$ & 17.7/27& & 0.66$_{-0.39}^{+0.52}$ & 0.38$_{-0.05}^{+0.07}$ & 0.84$_{-0.02}^{+0.02}$ & 1.4$_{-0.8}^{+0.7}$   & 22.3$_{-0.2}^{+0.2}$ & 0.32$_{-0.02}^{+0.02}$ & 26.0/47\\
				& LMC  & & 0.71$_{-0.23}^{+0.18}$ & 1.94$_{-0.15}^{+0.27}$  & 1.4$_{-0.1}^{+0.1}$ & 22.3$_{-0.2}^{+0.1}$ & 014$_{-0.10}^{+0.12}$ & 17.5/27  & & 0.52                              & 0.48                               & 0.84                               & 1.3                               &  23.0                           & 0.35                                & 122.9/47\\
				& MW   & & 0.76$_{-0.15}^{+0.15}$ & 2.29$_{-0.17}^{+0.18}$  & 1.4$_{-0.2}^{+0.1}$ & 22.3$_{-0.1}^{+0.2}$ & 0.10$_{-0.03}^{+0.06}$ & 17.7/27 & & 0.22                               & 0.57                               & 0.79                               & 0.8                               & 23.6                           & 0.15                                & 353.1/27\\
                & SMC$^{f}$ & & 0.40$_{-0.31}^{+0.35}$ & 0.93$_{-0.37}^{+0.72}$  & 1.4$_{-0.1}^{+0.1}$  & 22.1 & 0.30$_{-0.18}^{+0.09}$ & 19.8/28                 & & 0.66$_{-0.39}^{+0.53}$ & 0.36$_{-0.08}^{+0.08}$ & 0.84$_{-0.02}^{+0.02}$ & 1.4$_{-0.8}^{+0.8}$    & 22.1                            & 0.32$_{-0.02}^{+0.03}$ & 27.5/48\\
				& LMC$^{f}$ & & 0.66$_{-0.21}^{+0.20}$ & 1.95$_{-0.25}^{+0.29}$  & 1.4$_{-0.1}^{+0.1}$ & 22.1 & 0.17$_{-0.10}^{+0.10}$ &  20.1/28                  & & 0.58                               & 0.33                               & 0.88                               & 1.3                               &  22.1                           & 0.43                               & 118.8/48\\
				& MW$^{f}$  & & 0.72$_{-0.13}^{+0.17}$ & 2.35$_{-0.16}^{+0.16}$  & 1.4$_{-0.1}^{+0.1}$ & 22.1 & 0.11$_{-0.08}^{+0.06}$ & 20.2/28                   & & 0.0                                 & 0.29                               & 0.93                               & 1.1                               & 22.1                           & 0.42                                & 800.2/48\\               
               & XRT &$<$ 0.8 & 0.60$_{-0.16}^{+0.22}$ & & & & & 17.5/18 & & & & & & & & \\
               & SMC$^{p}$ & 0.52$_{-0.38}^{+0.50}$ & 0.40$_{-0.05}^{+0.07}$ & 0.78$_{-0.02}^{+0.02}$ & 0.9$_{-0.5}^{+3.4}$&                                   & 0.16$_{-0.05}^{+0.05}$ & 21.3/21 & & & & & & & &\\
               & LMC$^{p}$ & 0.52$_{-0.38}^{+0.51}$ & 0.47$_{-0.19}^{+0.31}$ & 0.78$_{-0.02}^{+0.02}$ & 1.0$_{-0.4}^{+4.1}$&                                   & 0.16$_{-0.06}^{+0.05}$ & 22.2/21 & & & & & & & &\\\vspace{0.25cm}
               & MW$^{p}$  & 0.48$_{-0.37}^{+0.50}$ & 0.76$_{-0.02}^{+0.02}$ & & &                                   & 0.10$_{-0.05}^{+0.05}$ & 20.0/22 & & & & & & & &\\

130427A & SMC  & & 0.43$_{-0.16}^{+0.10}$ & 0.79$_{-0.18}^{+0.20}$ & 0.7$_{-0.3}^{+0.3}$& & $<0.26$ & 31.7/58   & & 0.08$_{-0.02}^{+0.02}$ & 0.37$_{-0.04}^{+0.05}$ & 0.68$_{-0.01}^{+0.01}$ & 0.7$_{-0.2}^{+0.3}$ & & 0.16$_{-0.02}^{+0.02}$ & 129.3/147\\
               & LMC  & & 0.43$_{-0.19}^{+0.10}$ & 0.80$_{-0.16}^{+0.18}$ & 0.7$_{-0.3}^{+0.3}$& & $<0.29$ & 31.6/58   & & 0.11$_{-0.02}^{+0.02}$ & 0.05$_{-0.11}^{+0.10}$ & 0.77$_{-0.02}^{+0.03}$ & 1.0$_{-0.3}^{+0.7}$ & & 0.54$_{-0.09}^{+0.10}$ & 130.0/147\\
               & MW   & & 0.43$_{-0.20}^{+0.10}$ & 0.80$_{-0.29}^{+0.25}$ & 0.7$_{-0.3}^{+0.3}$& & $<0.21$ & 31.6/58   & & 0.12$_{-0.02}^{+0.02}$ & 0.03$_{-0.11}^{+0.11}$ & 0.81$_{-0.03}^{+0.03}$ & 1.3$_{-0.1}^{+0.6}$ & & 0.59$_{-0.10}^{+0.11}$ & 123.9/147\\
               & XRT &0.10$_{-0.03}^{+0.03}$ & 0.73$_{-0.09}^{+0.09}$ & & & & & 88.9/87 & & & & & & & & \\
               & SMC$^{p}$ & 0.10$_{-0.03}^{+0.04}$ & 0.33$_{-0.15}^{+0.14}$ & 0.73$_{-0.07}^{+0.10}$ & 1.6$_{-1.3}^{+7.8}$&                                  & 0.20$_{-0.12}^{+0.18}$ & 89.2/91 & & & & & & & &\\
               & LMC$^{p}$ & 0.10$_{-0.03}^{+0.04}$ & 0.31$_{-0.16}^{+0.17}$ & 0.73$_{-0.08}^{+0.10}$ & 1.4$_{-1.1}^{+4.8}$&                                   & 0.23$_{-0.19}^{+0.18}$ & 89.2/91 & & & & & & & &\\\vspace{0.25cm}
               & MW$^{p}$  & 0.10$_{-0.03}^{+0.03}$ & 0.31$_{-0.15}^{+0.15}$ & 0.73$_{-0.06}^{+0.09}$ & 1.4$_{-1.1}^{+5.8}$&                                   & 0.24$_{-0.19}^{+0.19}$ & 89.2/91 & & & & & & & &\\

130603B & SMC  & & 0.30$_{-0.16}^{+0.23}$ & 0.80$_{-0.31}^{+0.53}$ & 0.9$_{-0.2}^{+0.3}$ &  & 1.33$_{-0.38}^{+0.11}$ & 14.6/15 &  & 0.20$_{-0.09}^{+0.15}$ & 0.42$_{-0.22}^{+0.12}$ & 0.92$_{-0.04}^{+0.08}$ & 0.8$_{-0.1}^{+0.1}$ & & 1.19$_{-0.12}^{+0.23}$ & 21.3/23\\
			   & LMC  & & 0.24$_{-0.18}^{+0.25}$ & 0.86$_{-0.37}^{+0.53}$ & 0.7$_{-0.2}^{+0.2}$ &  &  1.40$_{-0.29}^{+0.25}$& 13.6/15 & &  0.21$_{-0.09}^{+0.16}$ & 0.30$_{-0.15}^{+0.11}$ & 0.95$_{-0.04}^{+0.05}$ & 0.7$_{-0.1}^{+0.1}$ & & 1.33$_{-0.12}^{+0.18}$ & 20.9/23\\
			   & MW   & & 0.24$_{-0.19}^{+0.24}$& 0.78$_{-0.32}^{+0.51}$ &  0.7$_{-0.2}^{+0.2}$&  &  1.45$_{-0.15}^{+0.35}$ & 14.3/15 & & 0.21$_{-0.10}^{+0.16}$ & 0.31$_{-0.14}^{+0.12}$ & 0.94$_{-0.04}^{+0.04}$ & 0.6$_{-0.1}^{+0.1}$ & & 1.36$_{-0.06}^{+0.16}$ & 21.2/23\\
               & XRT &$<$ 0.3 & 0.71$_{-0.35}^{+0.40}$ & & & & & 6.1/6 & & & & & & & & \\
               & SMC$^{p}$ & 0.17$_{-0.09}^{+0.13}$ & 0.31$_{-0.19}^{+0.21}$ & 0.86$_{-0.02}^{+0.03}$ & 0.5$_{-0.2}^{+1.5}$&                                  & 0.81$_{-0.07}^{+0.08}$ & 11.5/9 & & & & & & & &\\
			   & LMC$^{p}$ & 0.18$_{-0.09}^{+0.13}$ & 0.28$_{-0.26}^{+0.17}$ & 0.86$_{-0.02}^{+0.03}$ & 0.5$_{-0.2}^{+2.1}$&                                  & 0.83$_{-0.08}^{+0.10}$ & 11.1/9 & & & & & & & &\\ \vspace{0.25cm}
			   & MW $^{p}$ & 0.18$_{-0.09}^{+0.14}$ & 0.27$_{-0.22}^{+0.08}$ & 0.86$_{-0.02}^{+0.03}$ & 0.5$_{-0.2}^{+2.3}$&                                  & 0.88$_{-0.08}^{+0.08}$ & 10.6/9 & & & & & & & &\\

130606A & SMC & & 0.93 & & & 19.1 & 0.0 & 28.1/14 & &  < 3.5 & 0.96$_{-0.02}^{+0.02}$& & & 19.9$_{-0.2}^{+0.2}$ & < 0.01 & 48.8/31\\ 
				& LMC & & 0.94 & & & 19.1 & 0.0 & 28.6/14 & &  < 3.5 & 0.96$_{-0.02}^{+0.02}$& & & 19.9$_{-0.2}^{+0.2}$ & < 0.01 & 48.8/31\\
				& MW  & & 0.94 & & & 19.1 & 0.0 & 28.4/14 & &  < 3.5 & 0.96$_{-0.02}^{+0.02}$& & & 19.9$_{-0.2}^{+0.2}$ & < 0.01 & 48.8/31\\
				& SMC$^{f}$ & & 0.92 & & & 19.94 & 0.0 & 33.7/15 & &  < 3.4 & 0.96$_{-0.02}^{+0.02}$& & & 19.94 & < 0.01 & 50.1/32\\ 
				& LMC$^{f}$ & & 0.92 & & & 19.94 & 0.0 & 33.6/15 & &  < 3.4 & 0.96$_{-0.02}^{+0.02}$& & & 19.94 & < 0.01 & 50.1/32\\
				& MW$^{f}$  & & 0.93 & & & 19.94 & 0.0 & 33.8/15 & &  < 3.4 & 0.96$_{-0.02}^{+0.02}$& & & 19.94 & < 0.01 & 50.1/32\\
                & XRT &$<$ 2.2 & 0.77$_{-0.15}^{+0.20}$ & & & & & 13.3/15 & & & & & & & & \\
                & SMC$^{p}$ &< 3.6 & 0.96$_{-0.02}^{+0.03}$ & & &                                  & < 0.1 & 15.9/18 & & & & & & & &\\
			    & LMC$^{p}$ &< 3.6 & 0.96$_{-0.02}^{+0.04}$ & & &                                  & < 0.1 & 16.0/18 & & & & & & & &\\\vspace{0.25cm}
			    & MW$^{p}$  &< 3.6 & 0.96$_{-0.02}^{+0.07}$ & & &                                  & < 0.2 & 16.0/18 & & & & & & & &\\

\hline\hline
\end{longtable}
\begin{flushleft}
(p) Broadband fit using photometric measurements. \newline (f) Fit in which neutral hydrogen column density has been fixed to the value measured from the literature.
\end{flushleft}
\end{landscape}
\end{longtab}

\end{document}